\documentclass[floats,floatfix,amssymb,aps,prd,twocolumn,superscriptaddress,nofootinbib,preprintnumbers]{revtex4-2}
		
\usepackage{amssymb,amsmath,verbatim,mathtools,needspace,enumitem,etoolbox,graphicx,physics,microtype,afterpage,bm}

\usepackage[utf8]{inputenc}
\usepackage{amsmath,amssymb}
\usepackage{amsfonts}
\usepackage{bm}
\usepackage{courier}
\usepackage{graphics,orcidlink}
\usepackage{float}
\usepackage{amsmath}
\usepackage{amssymb}
\usepackage[normalem]{ulem} %
\usepackage[mathscr]{euscript}
\usepackage{acronym}
\usepackage{float}
\usepackage[caption=false]{subfig}
\usepackage{lipsum}
\usepackage{mathtools}
\usepackage{multirow}
\usepackage{graphicx}
\usepackage{mathtools}
\usepackage{multirow}
\usepackage{graphicx}
\usepackage{braket}
\usepackage{orcidlink}

\usepackage{ragged2e}
\DeclareCaptionJustification{justified}{\justifying}
\makeatletter
\newcommand{\subsetsim}{\mathrel{\mathpalette\subset@sim\relax}}
\newcommand{\subset@sim}[2]{%
  \vtop{\offinterlineskip\m@th
    \ialign{\hfil##\cr
      $#1\subset$\cr\noalign{\kern0.5pt}\scalebox{0.9}{$#1\sim$}\cr
    }%
  }%
}
\makeatother

\usepackage{amssymb,amsmath,verbatim,mathtools,needspace,enumitem,etoolbox,graphicx,physics,microtype,afterpage,bm}

\usepackage{multirow}
\usepackage{dcolumn}
\usepackage{pifont}
\usepackage{lmodern}

\usepackage{multirow}

\allowdisplaybreaks
\usepackage{tikz}
\usepackage{color}
\usepackage{framed}
\usepackage{hyperref}
\usepackage[all]{hypcap}
\definecolor{rossos}{cmyk}{0,1,1,0.55}
\definecolor{bluscuro}{rgb}{0.15, 0.2, .85}
\definecolor{bluchiaro}{cmyk}{1,.3,0.,0.1}
\definecolor{ForestGreen}{rgb}{0.13, 0.55, 0.13}
\definecolor{azure}{rgb}{0.0, 0.5, 1.0}
 
\usepackage{slashed}
\usepackage{hyperref}
\usepackage{tikz}
\usepackage[dvipsnames,svgnames,x11names]{xcolor}
\usetikzlibrary{arrows.meta,decorations.pathmorphing,decorations.markings}
\usetikzlibrary{calc}

\def\f{\frac}
\def\d{{\mathrm{d}}}

\newcommand{\odagg}{\mathcal{O}^{\dagger}}
\newcommand{\odaggone}{\mathcal{O}^{\dagger}_{(1)}}
\newcommand{\odaggzero}{\mathcal{O}^{\dagger}_{(0)}}
\newcommand{\hodagg}{\hat{\mathcal{O}}^\dagger}

\newcommand{\oone}{\mathcal{O}_{(1)}}
\newcommand{\ozero}{\mathcal{O}_{(0)}}
\newcommand{\Y}{\mathbf{Y}}

\newcommand {\psin}{\psi_{\boldsymbol{n}}}
\newcommand {\psinp}{\psi_{\boldsymbol{n}^{\prime}}}
\newcommand {\psinzero}{\psi^{(0)}_{\boldsymbol{n}}}
\newcommand {\psinpzero}{\psi^{(0)}_{\boldsymbol{n}^{\prime}}}
\newcommand {\psinone}{\psi^{(1)}_{\boldsymbol{n}}}

\newcommand {\omegan}{\omega_{\boldsymbol{n}}}
\newcommand {\omeganp}{\omega_{\boldsymbol{n}^{\prime}}}

\newcommand{\surf}{\int_{\Sigma_t} \mathrm{d}\Sigma_a}

\newcommand{\normQNM}{\langle\! \langle \psi^{(0)}_{\boldsymbol{n}}, \psi^{(0)}_{\boldsymbol{n}} \rangle \!\rangle}
\newcommand{\projnone}{\langle\! \langle \psi^{(0)}_{\boldsymbol{n}}, \psi^{(1)}_{\boldsymbol n} \rangle \!\rangle}

\newcommand{\normp}{\langle\! \langle \psi^{(0)}_{\boldsymbol n^{\prime}}, \psi^{(0)}_{\boldsymbol n^{\prime}} \rangle\! \rangle}

\newcommand{\projnpone}{\langle \!\langle \psi^{(0)}_{\boldsymbol n^{\prime}}, \psi^{(1)}_{\boldsymbol n} \rangle \!\rangle}

\newcommand{\ps}{\Psi_2^{4/3} \mathcal{J}}

\newcommand{\be}{\begin{equation}}
\newcommand{\ee}{\end{equation}}
\renewcommand{\d}{{\rm d}}

\newcommand{\lp}{\left (}
\newcommand{\rp}{\right )}

\def\lsim{\mathrel{\rlap{\lower4pt\hbox{\hskip0.5pt$\sim$}}
    \raise1pt\hbox{$<$}}}         %
\def\gsim{\mathrel{\rlap{\lower4pt\hbox{\hskip0.5pt$\sim$}}
    \raise1pt\hbox{$>$}}}         %

\makeatletter
\def\l@subsubsection#1#2{}
\makeatother

\newcommand{\nottingham}{Nottingham Centre of Gravity \& School of Mathematical Sciences, University of Nottingham, University Park, Nottingham, NG7 2RD, UK}

\begin{document}
\title{
Schr\"odinger perturbation theory for black hole quasinormal modes
}

\begin{abstract}

Deviations from vacuum general relativity (such as modified theories or the presence of an environment) produce small shifts in black hole quasinormal mode (QNM) spectra. These effects are becoming increasingly relevant for gravitational wave astronomy as observations of ringdown spectra become more precise.
To leading order in deviations from vacuum general relativity, frequency shifts can be calculated using eigenvalue perturbation theory. However, for Kerr, there is no systematic framework to compute higher order corrections.
The major obstacle is that QNMs do not form a complete basis due to the non-self-adjointness of the system. Nevertheless, it was recently shown that QNMs are  orthogonal with respect to an appropriate bilinear form. In this work, we use the bilinear form to systematically lift Schr\"odinger perturbation theory to the black hole setting. We obtain a formula for quasinormal frequency shifts to any order, in terms of lower order mode shifts. We also provide a spectral decomposition of the first-order mode shift, which involves projections onto unperturbed QNMs along with continuous-spectrum contributions---making incompleteness explicit. We illustrate the framework on slowly-spinning Kerr and P\"oschl-Teller examples, where we find that the QNM sum itself diverges.

\end{abstract}

\author{Jacopo Lestingi\,\orcidlink{0009-0009-4366-7896}}
 \email{jacopo.lestingi@nottingham.ac.uk}
\affiliation{\nottingham}

\author{Laura Sberna}
\affiliation{\nottingham}

\author{Stephen R. Green\,\orcidlink{0000-0002-6987-6313}}
\affiliation{\nottingham}

\date{\today}
\maketitle

\section{Introduction}\label{sec:intro}

In vacuum general relativity (GR), the quasinormal mode (QNM) spectrum of a Kerr black hole (BH) is entirely characterized by its mass $M$ and spin $a$~\cite{Nollert:1999ji, Kokkotas:1999bd, Berti:2009kk}. This universality underlies the BH spectroscopy program, which uses quasinormal frequencies
extracted from gravitational wave data to test the no-hair theorem and search for new physics. Deviations from the vacuum-GR spectrum could indicate the presence of beyond-GR (bGR) physics or of an astrophysical environment surrounding the BH remnant~\cite{Berti:2025hly}. As the sensitivity of gravitational wave observations improves, deviations from Kerr will eventually become observationally accessible. Accurately predicting these frequency shifts has therefore become an important problem in BH perturbation theory.

Many modified theories of gravity are continuously connected to GR through a small coupling parameter $\zeta$. These theories can affect the QNM spectrum via two main channels. First, the linearized gravitational field equation about the stationary BH solution differs from that of GR. This can lead to $\zeta$-dependent shifts away from the Kerr quasinormal frequencies (see, e.g.,~\cite{Cano:2024ezp,Cano:2023jbk, Chung:2023wkd, Chung:2024ira, Chung:2024vaf, Chung:2025gyg, Li:2025fci, Aly:2026otj, Blazquez-Salcedo:2024dur,Boyce:2026rnn}). Second, the modified theory may contain new dynamical fields non-minimally coupled to gravity. Although these are not observed directly, they can imprint their characteristic frequencies on those of the metric~\cite{Crescimbeni:2024sam,Lestingi:2025jyb,Crescimbeni:2025kxi}. The new fields can also mediate shifts in the gravitational quasinormal frequencies~\cite{Li:2023ulk}.
These effects are naturally captured via modified Teukolsky equations~\cite{Li:2022pcy,Cano:2023tmv} (which have also been used in the context of accretion disks and boson clouds~\cite{Dyson:2025dlj, Dyson:2026ddd, Li:2025ffh}).

Perturbative approaches to compute spectral shifts at first order in the coupling parameter have been developed over the past years. For modifications of Regge-Wheeler-like equations, Hatsuda and Kimura \cite{Hatsuda:2019eoj,Hatsuda:2023geo} developed a framework to compute QNM spectral shifts based on directly solving perturbed versions of the master equation and boundary conditions. 
A different approach was developed for Kerr, first for 
specific deformations~\cite{Mark:2014aja,Zimmerman:2014aha}, and then for the general modified Teukolsky equation~\cite{Hussain:2022ins,Li:2023ulk} (see also \cite{Li:2025fci,Cano:2024ezp,Cano:2024jkd,Weller:2024qvo,Wu:2025obg,Aly:2026otj} for recent applications). This approach used an eigenvalue perturbation theory, which is based on a product on QNMs derived from Sturm-Liouville theory~\cite{Berti:2025hly}. The product involves a complex integration such that it is finite on QNMs (despite their exponential growth at the bifurcation surface and spatial infinity), resulting in a formula for the first-order frequency shift that resembles that of quantum mechanical perturbation theory. However, the product used in these studies is not orthogonal on Kerr QNMs, so it is not clear how to write down formulas at higher order. 
A systematic perturbative framework for computing higher order Kerr QNM spectral shifts due to generic perturbations of the Teukolsky operator is therefore currently not available.

We describe such a framework here, using a recently-developed bilinear form on Kerr, under which QNMs are orthogonal~\cite{Green:2022htq}. The bilinear form is based on the symplectic current and is automatically conserved on solutions to the Teukolsky equation. The associated conservation law implies QNM orthogonality and underpins the systematic construction of a Schr\"odinger-style perturbative expansion. 

Our framework gives rise to a generic formula for the frequency shifts at any order in the perturbative parameter. At first order, this reduces to the known frequency-shift formula of eigenvalue perturbation theory. Starting from second order, the frequency shifts depend on the lower-order mode shifts. However, Kerr QNMs are known not to form a complete basis \cite{Leaver:1986gd,Arnaudo:2025uos,Su:2026fvj,Warnick:2022hnc} (as the problem is non-Hermitian, with energy dissipated through the horizon and away to infinity), so it is not possible to directly express these mode shifts as a sum of only background QNMs.

Using Green's function techniques, we derive a spectral decomposition of the mode shift. This features a sum over background QNMs and a continuous spectral component---an integral along the branch cut and the high-frequency arc in the complex frequency plane. The continuous term is analogous to the contribution of the unbound spectrum in quantum mechanical systems such as the hydrogen atom. Unlike in quantum mechanics, however, no spectral theorem stands behind the QNM expansion, and the decomposition remains formal. 
We then show that the spectral decomposition can be rewritten in terms of projections using the bilinear form. This reveals that the QNM component contains two types of projections: the projection of the mode shift onto background modes, mirroring Schrödinger perturbation theory, and the projection of a secular term on background modes, absent in quantum mechanics.

We apply our framework to two examples. First, we consider the  P\"oschl-Teller potential (where the mode shift is known analytically), with its width as the perturbative parameter. In this case, we show that the QNM contribution alone to the first order mode shift diverges. %
As the mode shift is a finite quantity, this suggests that it may be possible to regularize the QNM contribution using the continuum piece. However, as this is challenging to evaluate, we use instead the  known analytic expression for the mode shift in the formula for the second-order frequency shift, and we find agreement with the known result. We also validate the second-order frequency-shift formula using slowly-spinning Kerr (using the first-order mode shift from the Leaver series). Our work builds upon Ref.~\cite{Cannizzaro:2023jle}, which used the same bilinear form to calculate first-order scalar frequency shifts for BH boson clouds around Kerr. Similar techniques were also used to explore turbulence in the vicinity of near-extreme Kerr~\cite{Iuliano:2024ogr}, and the emission of QNMs by BH boson clouds \cite{Cannizzaro:2025vpb}.

Our paper is organized as follows. In Sec.~\ref{sec: Product}, we summarize the derivation and main properties of the bilinear form. We present the Schr\"odinger-style QNM perturbation theory in Sec.~\ref{sec: Formalism}, deriving the formula to compute the frequency shift at any order. In Sec.~\ref{sec:Modeshifts}, we describe the computation of the mode shift along with the obstacle posed by QNM incompleteness. We apply the formalism to the toy problem of P{\"o}schl-Teller in Sec.~\ref{sec: PT potential}, and to slowly-spinning Kerr in Sec.~\ref{sec: SchwToKerr}. We conclude in Sec.~\ref{sec:conclusions}.

\section{Preliminaries}\label{sec: Product}

\subsection{Green's function and quasinormal modes}
\label{sec:gf}

For Kerr perturbations, the Teukolsky formalism reduces the solution of the linearized Einstein equation to that of a complex scalar wave equation~\cite{Teukolsky:1973ha},
\begin{equation}
  \label{eq: sec A - Teukolsky equation}
  {}_s\mathcal{O} \psi = 0,
\end{equation}
where the spin $s=\pm 2$. Here $\psi$ represents an extreme Weyl scalar perturbation, either $\psi_0$ (for $s=2$) or $\Psi_2^{-4/3}\psi_4$ (for $s=-2$), where $\Psi_2$ is the background Weyl scalar.

In  Boyer-Lindquist (BL) coordinates $(t,r,\theta,\phi)$, the Teukolsky equation written as ${}_s T = \Sigma \, {}_s\mathcal{O}$ with $\Sigma= r^2 +a^2 \cos^2\theta$ is separable \cite{Teukolsky:1973ha}. For an initial-value problem, we pass to the frequency domain using the Laplace transform,
and decompose over the BL spatial coordinates. The time-domain field is recovered by the inversion integral,
\begin{equation}
    \label{eq: wave function decomposition}
    \psi = \frac{1}{2\pi}\sum_{\ell \ge 2}\sum_{m=-\ell}^\ell \int_{-\infty+i\epsilon}^{\infty+i\epsilon} \mathrm{d}\omega\, e^{-i \omega t + i m \phi} {}_sS_{\ell m \omega}(\theta){}_sR_{\ell m \omega}(r).
\end{equation}
The angular and radial functions satisfy separated ordinary differential equations. Regular eigensolutions ${}_sS_{\ell m\omega}$ to the homogeneous angular equation are called spin-weighted spheroidal harmonics, and are indexed by $\ell\ge |s|$. They depend on the BH spin $a$ only via the combination $a\omega$. The radial equation has a source corresponding to initial data. It also couples to the angular equation via a separation constant, which is determined once the angular solutions are fixed.

The radial equation must be solved subject to appropriate (dissipative) boundary conditions, and the standard approach is to use Green's function methods. Let ${}_sR^\text{in}_{\ell m \omega}$, ${}_sR^\text{up}_{\ell m \omega}$ be homogeneous solutions ingoing at the horizon and outgoing at infinity, respectively. Then the radial Green's function reads
\begin{equation}\label{eq:GF_radial_main}
    {}_sg_{\ell m\omega}(r,r') = - \Delta^s(r') \frac{{}_sR^{\text{in}}_{\ell m\omega}(r_<){}_sR^{\text{up}}_{\ell m\omega}(r_>)}{{}_sW_{\ell m}(\omega)},
\end{equation}
where $\Delta = r^2 -2 Mr + a^2$ and $r_>$ ($r_<$) is the greater (lesser) of $r$ and $r'$. We also defined the $\Delta$-scaled Wronskian as
\begin{equation}
    W[R^{\text{in}},R^{\text{up}}] = \Delta^{1+s} \lp R^{\text{in}} \frac{\d R^{\text{up}}}{\d r} - R^\text{up} \frac{\d R^\text{in}}{\d r} \rp.
\end{equation}
The Wronskian is constant in $r$ for $R^\text{in}, R^\text{up}$ solutions of the homogeneous radial Teukolsky equation, so we write $W=W(\omega)$. The inhomogeneous solution ${}_sR_{\ell m \omega}(r)$ can then be expressed as an integral of the Green's function against the initial data \cite{Berti:2006wq}.

To evaluate the $\omega$ integral in \eqref{eq: wave function decomposition} to return to the time domain, we deform the contour into the lower half of the complex-$\omega$ plane. This gives rise in the standard way to an integral along a branch cut, an arc at infinity, and residues at a discrete set of poles. The poles are due to the zeroes of the Wronskian, and define the complex quasinormal frequencies~\cite{Kokkotas:1999bd,Berti:2006wq}. For each angular harmonic, we label the tower of these frequencies by their overtone number $n$ and mirror index $p=\pm1$, denoting the frequency as $\omega_{\boldsymbol n}(M,a)$, where the multi-index $\boldsymbol n \equiv p \ell m n$. Since the Wronskian vanishes at QNM frequencies, the two radial functions become linearly dependent, $R_{\boldsymbol n}^\text{in}= A_{\boldsymbol n}R_{\boldsymbol n}^\text{up}$. The common solution satisfies both the ingoing at horizon and outgoing at infinity boundary conditions and is the QNM radial function. We denote the full QNM wavefunction as $\psi_{\boldsymbol n}(t,r,\theta,\phi)$.

The quasinormal spectrum can be computed using, for instance, the continued fraction method of Leaver~\cite{Leaver:1985ax}. The spectrum is entirely determined by the BH mass and spin, so the gravitational-wave observation of more than one ringdown mode gives rise to a consistency condition; this underpins the BH spectroscopy program to test general relativity. Although $(M,a)$ determine the frequencies, the amplitude of an individual mode is fixed by initial data and is given by the residue at the corresponding pole (the \emph{excitation coefficient}, which we return to in Sec.~\ref{subsec: excitation coefficients}).

We emphasize that the frequency enters this problem nonlinearly: $\omega$ appears in the radial equation both quadratically and linearly, with $r$-dependent coefficients, as well as through the spheroidal coupling $a\omega$ and the separation constant. The QNM problem is a nonlinear eigenvalue problem, with quasinormal frequencies arising as zeroes of $W(\omega)$, rather than as eigenvalues of a fixed operator. By contrast, the Schr\"odinger problem $(H-E)\ket{\psi} = 0$ is linear in $E$. This nonlinearity will affect the mode expansions of Sec.~\ref{sec:Modeshifts}.%

Although QNMs describe the BH ringdown for much of its evolution, they do not in general form a complete solution to the initial value problem. Indeed QNMs are resonances of an open, dissipative system: its frequencies are complex and there is no spectral theorem to guarantee completeness, or convergence of mode expansions, in contrast to the bound states of a self-adjoint system. At the practical level, completeness is spoiled by the continuous branch cut and arc contributions to the inverse Laplace transform. The branch point is at $\omega=0$, and arises from the properties of the radial functions $R^\text{in}$, $R^\text{up}$. The associated cut is usually placed along the negative imaginary axis, and is responsible for both the polynomial decay of the ringdown at late times~\cite{Leaver:1986gd} as well as an initial direct burst of radiation \cite{Su:2026fvj}. 
The QNM expansion is not globally valid, rather only after a certain amount of time has passed~\cite{Leaver:1986gd}. Precise conditions for the validity of this expansion, as well as associated prescriptions for closing the complex-$\omega$ contour, were recently worked out in detail~\cite{DeAmicis:2025xuh,Kuntz:2025gdq,Arnaudo:2025uos,Su:2026fvj,Arnaudo:2026tcy,DeAmicis:2026wqd}. QNM incompleteness also represents an obstacle to Schr\"odinger-style perturbation theory beyond first-order frequency shifts, which we describe in Sec.~\ref{sec:Modeshifts}, although frequency shifts can still be computed at all orders. %

\subsection{Bilinear form} \label{subsec: defining the product}

Although QNMs are incomplete, they are nevertheless orthogonal under a suitable bilinear form. This was shown for Kerr in \cite{Green:2022htq}, building on prior work of Leung and collaborators~\cite{PhysRevA.49.3057,Leung:1997was}. The QNM orthogonality relation will underpin our Schr\"odinger perturbation framework, so we summarize it here. The bilinear form is based on the symplectic current for the Teukolsky equation. Its construction is therefore very much analogous to that of the Klein-Gordon product. However, rather than using complex conjugation, we use $t$--$\phi$ reflection (a symmetry of the Teukolsky equation) to conjugate solutions.

From now on we fix $\mathcal O \equiv {}_2\mathcal O$ to be the spin $s=2$ Teukolsky operator. Given a pair of Geroch-Held-Penrose (GHP) scalars $\psi$, $\tilde\psi$, of spin weights $-2$, $+2$, respectively, we define the adjoint operator $\mathcal O^\dagger$ such that
\begin{equation}
  \label{symplectic current}
  \psi \mathcal{O} \tilde{\psi} - \tilde{\psi} \odagg \psi = \nabla_a \pi^a[\tilde{\psi} ,\psi].
\end{equation}
The adjoint can be evaluated using an integration-by-parts-like application of the Leibniz rule, and it is easily verified that $\mathcal O^\dagger = {}_{-2}\mathcal O$ is the $s=-2$ Teukolsky operator. A total divergence arises as a residual on the right-hand side. The  vector field $\pi^a$ is known as the symplectic current.

To obtain a bilinear form, we integrate the symplectic current on a constant time slice $\Sigma_t$ of Kerr,
\begin{equation}
  \label{bilinear form}
  \Pi[\tilde{\psi},\psi] \equiv \int_{\Sigma_t} \mathrm{d}\Sigma_a\, \pi^a[\tilde{\psi},\psi] .
\end{equation}
For this to be well-defined, we assume for now that $\psi$, $\tilde\psi$ have compact spatial support. The $\Pi$ product takes as arguments two GHP scalars of opposite spin weights, and it is linear in both, hence it is known as a bilinear form. $\Pi$ is conserved on solutions, in the sense that it is invariant under deformations of $\Sigma_t$. %

To obtain a bilinear form that takes two scalars of the \emph{same} weight, we use the discrete $t$--$\phi$ reflection isometry of Kerr. Indeed, one can define a GHP-covariant operator $\mathcal J$, which corresponds to the mapping $(t,\phi)\to(-t,-\phi)$ followed by a GHP prime operation~\cite{Green:2022htq}. It therefore takes $s = -2$ weighted fields into $s=+2$. Moreover, $\mathcal J$ satisfies the intertwining property,
\begin{equation}
  \mathcal{O} \Psi_2^{4/3} \mathcal{J} = \Psi_2^{4/3} \mathcal{J} \odagg.
\end{equation}
Thus it takes $\ker \odagg$ into $\ker \mathcal O$. In BL coordinates and the Kinnersley frame,
\begin{equation} 
  \label{reflection in BL coordinates}
  \Psi_2^{4/3} \mathcal{J} \psi = 4 M^{4/3} \Delta^{-2} \left. \psi \right|_{(t,\phi)\rightarrow (-t,-\phi)},
\end{equation} 
where $\Delta = r^2 + a^2 -2 M r$.

Finally, we define the bilinear form on two compact support $s=-2$ scalars as
\begin{equation}
  \label{eq: Product}
  \langle\! \langle \psi_1 , \psi_2 \rangle \!\rangle = \int_{\Sigma_t} \mathrm{d}\Sigma_a\, \pi^a[ \Psi_2^{4/3} \mathcal{J} \psi_1,\psi_2].
\end{equation}
In BL coordinates and the Kinnersley frame, this reads
\begin{widetext}
  \begin{eqnarray}
    \label{eq: the product in coordinate}
    \langle\!\langle \psi_1, \psi_2 \rangle\!\rangle & =& -4M^{4/3} \int_{r_+}^{\infty} \int_0^{\pi} \int_0^{2 \pi} \mathrm{d}r \, \mathrm{d}\theta \, \mathrm{d}\phi \, \frac{\sin\theta}{\Delta^2} 
\left[ 
\psi_1 \Bigg|_{\substack{t \to -t \\ \phi \to -\phi}}
\left( \frac{\Lambda}{\Delta} \partial_t + \frac{2Mra}{\Delta} \partial_\phi + 2 \left[ -r - i a \cos\theta + \frac{M}{\Delta} (r^2 - a^2) \right] \right) \psi_2 
\right.
 \nonumber \\ &&
\left.
\qquad \qquad
+ \psi_2 \left[
\left( \frac{\Lambda}{\Delta} \partial_t + \frac{2Mra}{\Delta} \partial_\phi + 2 \left[ -r - i a \cos\theta + \frac{M}{\Delta} (r^2 - a^2) \right] \right) \psi_1 \right]
\Bigg|_{\substack{t \to -t \\ \phi \to -\phi}}
\right],
    \end{eqnarray}
\end{widetext}
where $\Lambda = (r^2 + a^2)^2 - \Delta a^2 \sin^2 \theta$. Note that the bilinear form contains time derivatives: it involves full Cauchy data $(\psi, \partial_t \psi)$ on $\Sigma_t$, so for a QNM, the frequency is part of these data, $\partial_t \psin = -i \omegan \psin$.  %

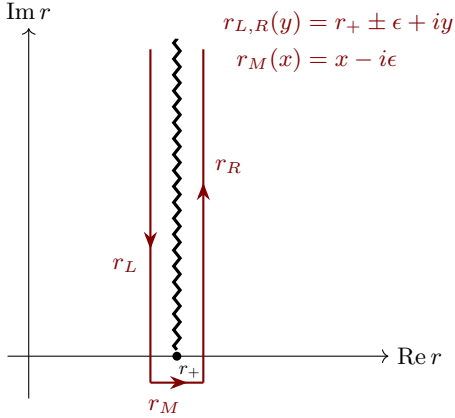
\begin{figure}[t]
  \tikzset{
  midarrow/.style={
    postaction={decorate},
    decoration={markings, mark=at position 0.6 with {\arrow{Stealth}}}
  }
}

\tikzset{
  midarrowcircle/.style={
    postaction={decorate},
    decoration={markings, mark=at position 0.7 with {\arrow{Stealth}}}
  }
}

\begin{tikzpicture}[scale=1.4
]
  \def\rHx{1.4}   %
  \def\Y{3.0}     %
  \def\eps{0.25}  %

  \draw[->] (-0.2,0) -- (3.4,0) node[right] {$\Re r$};
  \draw[->] (0,-0.4) -- (0,3.1) node[above] {$\Im r$};

  \coordinate (rH) at (\rHx,0);
  \fill (rH) circle (1.2pt) ;
  \node[below right, xshift=-2.5pt, yshift=-0.2pt] at (rH) {$\scalebox{0.8}{$r_+$}$};

\draw[very thick, decorate,
      decoration={zigzag, segment length=6pt, amplitude=1.2pt}]
  ($(rH)+(0,0.06)$) -- ++(0,\Y-0.06);

  \draw[rossos, thick, midarrow]
    (\rHx+\eps,-\eps) -- (\rHx+\eps,\Y-0.1)
    node[above right, xshift=4pt, yshift=2pt] {$r_{L,R}(y)=r_+ \pm \epsilon+i y$}
    node[above right, xshift=1pt, yshift=-50pt] {$r_{R}$}
    node[above right, xshift=9pt, yshift=-12pt] {$r_{M}(x)=x-i\epsilon$};

  \draw[rossos, thick, midarrow]
    (\rHx-\eps,\Y-0.1) -- (\rHx-\eps,-\eps)
    node[below left, xshift=-1pt, yshift=50pt] {$r_{L}$};

  \draw[rossos, thick, midarrowcircle]
(\rHx-\eps,\Y-3-\eps) -- (\rHx+\eps,\Y-3-\eps)
node[below left, xshift=-5pt, yshift=-3pt] {$r_{M}$};

\end{tikzpicture}
\caption{Integration contour $\mathcal{C}$ used to regularize radial integrals featuring two QNMs $\psin, \psinp$, as is the case for the bilinear form on QNMs, when ${\rm Re}(\omegan+\omeganp)>0$. When ${\rm Re}(\omegan+\omeganp)<0$ one can simply reflect both the branch cut and the contour with respect to the horizontal axis.
The contour sections $r_M$ and $r_{L,R}$ are parametrized by $x \in [r_+ -\epsilon,r_++\epsilon]$ and $y \in [-\epsilon,+\infty)$, respectively, with $\epsilon \ll 1$. The radial part of QNMs decays exponentially on $r_{L,R}$, making the radial integration finite. 
}
\label{plot: Leaver's contour}
\end{figure}

QNM wavefunctions are well-known to diverge at the bifurcation surface and spatial infinity for $\Im \omega < 0$ due to their asymptotic in-/up- going boundary conditions. Hence, the definition of the bilinear form for fields of compact support would give divergent integrals if applied directly to QNMs. To extend the definition to QNMs, we deform the radial integration into the complex-$r$ plane, defining a complex slice $\Sigma_\mathbb{C}$. The complex contour is shown in Fig.~\ref{plot: Leaver's contour}, and is chosen so that the QNMs decay exponentially to zero at both ends of the contour \cite{Leaver:1986gd,Ma:2024qcv,Berti:2025hly,Cannizzaro:2025vpb,Minucci:2026dgo}. Under this extended bilinear form, QNMs of different frequency will turn out to be orthogonal. %

\subsection{Conservation and mode orthogonality} \label{subsec: QNM orthogonality}

Given $\psi_1$, $\psi_2$ of compact support on $\Sigma_t$ (or decaying sufficiently rapidly at its endpoints), it follows from Cartan's magic formula that the time derivative of the bilinear form is given by \cite{Cannizzaro:2023jle} 
\begin{equation}
\label{Master Equation}
     \frac{\mathrm{d} }{\mathrm{d}t} \langle\! \langle \psi_1 , \psi_2 \rangle \!\rangle = \int_{\Sigma_t} \mathrm{d}\Sigma_a t^a\, \nabla_b\pi^b[ \Psi_2^{4/3} \mathcal{J} \psi_1,\psi_2],
\end{equation}
where $t^a$ is the time Killing vector field of the Kerr spacetime, and where in BL coordinates $\mathrm{d}\Sigma_a t^a = \mathrm{d}r\mathrm{d}\theta\mathrm{d}\phi \Sigma \sin\theta$.
This balance law holds regardless of whether $\psi_{1,2}$ are solutions, hence it is useful in perturbative calculations and will form the basis of the framework we develop here. Although QNMs do not decay on a real slice, they do on the complex slice $\Sigma_\mathbb{C}$, so the balance law extends to them as well. Eq.~\eqref{Master Equation} has been used in \cite{Cannizzaro:2023jle} in the context of $s=0$ fields to compute first-order frequency shifts for scalar clouds around rotating BHs due to self-gravity, and in \cite{Iuliano:2024ogr} to study nonlinearities in extremal Kerr BHs. %

If $\psi_{1,2} \in \ker \odagg$, then  $\nabla_a \pi^a[\Psi_2^{4/3} \mathcal{J} \psi_1,\psi_2] = 0$ by definition \eqref{symplectic current}. Thus, the bilinear form is conserved on-shell, $\frac{\mathrm{d} }{\mathrm{d}t} \langle \!\langle \psi_1 , \psi_2 \rangle \!\rangle = 0$. Fixing two QNM solutions $\psin$, $\psinp$, we therefore have
\begin{equation}
    0 = \frac{\mathrm{d} }{\mathrm{d}t} \langle \!\langle \psinp , \psin \rangle \!\rangle = -i (\omegan - \omeganp) \langle \!\langle \psinp , \psin \rangle \!\rangle \, .
\end{equation}
Hence, either $\omeganp = \omegan$ or $\langle \!\langle \psinp , \psin \rangle \!\rangle = 0$. QNMs with different frequencies are orthogonal to each other. 

For $\omeganp = \omegan$, we obtain the QNM \emph{norm} $\langle \!\langle \psin , \psin \rangle \!\rangle$. As the bilinear form is not a positive-definite inner product, the norm need not be positive and is in general complex valued. As we shall see, it can be interpreted as an excitation factor.

\subsection{Excitation of QNMs}\label{subsec: excitation coefficients}

Consider the equation $\odagg \psi = 0$, subject to some initial data specified on a Cauchy hypersurface $\Sigma_{t_0}$. It is natural to ask whether the bilinear form can be used as an orthogonal projector to directly obtain the amplitude of QNMs in a mode expansion. If that were the case, then during the period in which QNMs dominate the evolution, we could write
\begin{equation}
  \psi \simeq \sum_{\boldsymbol n} c_{\boldsymbol n} \psin ,
\end{equation}
with  
\begin{equation}\label{eq:cn-bilinear}
    c_{\boldsymbol n} = \frac{\langle \! \langle \psin, \left.\psi\right|_{t_0} \rangle \! \rangle}{\langle \! \langle \psin, \psin \rangle \! \rangle} 
\end{equation}
known as the excitation coefficient \cite{Berti:2006wq}.

There is already a well-known expression for excitation coefficients, which is obtained using Laplace transform methods:
\begin{equation}\label{eq:cn-laplace}
  c_{\boldsymbol n} = - \frac{i A_{\boldsymbol n}}{ \left.\frac{\d W}{\d \omega}\right |_{\omegan} }\int_{r_+}^{\infty} \d r' \Delta^{-2} I_{\boldsymbol n}(r')R_{\boldsymbol n}(r'). 
\end{equation}
Here, $I_{\boldsymbol n}(r')$ is a projection of the initial data onto appropriate angular modes, and involves the QNM frequency $\omega_{\boldsymbol n}$ \cite{Berti:2006wq}. %
Recall $A_{\boldsymbol{n}}$ relates the homogeneous radial solutions at a quasinormal frequency, $R^\text{in}(\omegan) = A_{\boldsymbol{n}} R^\text{up}(\omegan)$.

Although it is not immediately apparent, it was shown in \cite{Green:2022htq} (following the approach of \cite{PhysRevA.49.3057}) that \eqref{eq:cn-bilinear} and \eqref{eq:cn-laplace} are in full agreement.
To see this, first consider the coordinate bilinear form \eqref{eq: the product in coordinate}. It is straightforward to verify that the projector $\langle \! \langle \psin, \left.\psi\right|_{t_0} \rangle \! \rangle$ in the numerator of \eqref{eq:cn-bilinear} gives rise to the integral in \eqref{eq:cn-laplace}. In particular, it involves $\psi|_{t_0}$, its time derivative $\partial_t\psi|_{t_0}$, and $\omega_{\boldsymbol n}$. For the norm of the QNM appearing in the denominator of \eqref{eq:cn-bilinear}, a very nontrivial calculation shows that it is related to the derivative of the Wronskian \cite{Green:2022htq},
\begin{equation}
  \label{excitation factor - Wronskian derivative}
  \langle \! \langle \psin, \psin \rangle \! \rangle = \frac{4iM^{4/3}}{A_{\boldsymbol n}} \left.\frac{\mathrm{d} W}{\mathrm{d}\omega} \right|_{\omegan}.
\end{equation}
With this identification, the two expressions for the excitation coefficient coincide. 

\subsection{Sturm-Liouville product}

Here, we introduce the Sturm-Liouville (SL) product, which will be useful for drawing connections to quantum mechanics, as well as past work on eigenvalue perturbation theory. It will also allow us to express the QNM norm in terms of the frequency-domain Teukolsky operator. The SL product is defined as
\begin{equation}
  \langle \cdot,\cdot \rangle \equiv \surf t^a (\ps \psi_1)\, \psi_2.
\end{equation}
When acting on QNMs, the radial integration is along the complex contour of Fig.~\ref{plot: Leaver's contour}, so it remains finite. The SL product reduces to the frequency-domain product of the eigenvalue perturbation theory of Refs.~\cite{Mark:2014aja,Zimmerman:2014aha,Hussain:2022ins,Li:2023ulk,Berti:2025hly} when acting on functions of the form $\psi=e^{i m \phi}e^{-i \omega t}f(r,\theta)$, see also \cite{London:2023aeo}.

The SL product is defined with an appropriate weight based on the Sturm-Liouville form of the radial and angular Teukolsky equations, such that these equations are self-adjoint with respect to the product. Specifically, consider two mode functions $\psi_{\omega_1}=e^{i m_1 \phi}e^{-i \omega_1 t}f(r,\theta)$, $\psi_{\omega_2}=e^{i m_2 \phi}e^{-i \omega_2 t}g(r,\theta)$ (not necessarily solutions). For any $\omega$, one has the self-adjoint property $\int \mathrm{d}\phi\mathrm{d}r\mathrm{d}\theta \sin\theta \Delta^{-2}(\psi_{\omega_1} \hat T_{\omega}\psi_{\omega_2} - \psi_{\omega_2}\hat T_{\omega}\psi_{\omega_1}) = 0$, where $\hat T_{\omega}$ is the frequency-domain Teukolsky operator. This property was first exploited in Refs.~\cite{Mark:2014aja,Zimmerman:2014aha,Hussain:2022ins,Li:2023ulk}, and underlies their eigenvalue perturbation theory. Here, we use a hat on top of a time-translation-invariant operator $A$, to denote its frequency-domain form, defined by $A(e^{-i\omega t}\chi) = e^{-i\omega t}\hat{A}_{\omega}\chi = \hat{A}_{\omega}(e^{-i\omega t}\chi)$.

In order to re-express the norm in terms of the SL product, we start from the identity
\begin{equation}
  \label{eq: odagg-tpsin}
  \odagg(-i t \psin) = \partial_\omega \hodagg\big|_{\omegan} \psin .
\end{equation}
This follows by writing $\psin = e^{-i\omegan t}\chi_{\boldsymbol n}$ and using $te^{-i\omega t} = i \partial_\omega e^{-i\omega t}$ together with $\hodagg_{\omegan}\chi_{\boldsymbol n} = 0$. Consider now $\langle\!\langle \psin, -it \psin \rangle\!\rangle$. On the one hand, since the bilinear form \eqref{eq: the product in coordinate} is first order in time derivatives, this equals $-it \langle\!\langle \psin, \psin\rangle\!\rangle$ plus a constant, so its time derivative is $-i \langle\!\langle \psin, \psin\rangle\!\rangle$. On the other hand, the balance law \eqref{Master Equation} gives for the same derivative $-\surf t^a (\ps\psin) \odagg(-it\,\psin)$, since $\ps\psin$ solves the Teukolsky equation. Equating the two and using \eqref{eq: odagg-tpsin}, we obtain
\begin{equation}
    \label{eq: norm as operator derivative}
  \langle \! \langle \psin, \psin \rangle \! \rangle = -i \left\langle \psin, \partial_\omega \hodagg\big|_{\omegan} \psin \right\rangle.
\end{equation}

Eq.~\eqref{eq: norm as operator derivative} relates the bilinear form and the SL product on the diagonal. We now proceed to study the relation off the diagonal, where the bilinear form gives orthogonality. Consider the balance law \eqref{Master Equation} on the generic mode functions $\psi_{\omega_1}$, $\psi_{\omega_2}$. On the left side, we have $-i (\omega_2 - \omega_1) \langle \! \langle \psi_{\omega_1},\psi_{\omega_2} \rangle \! \rangle$. On the right side, using the intertwining property, we find $\surf t^a (\psi_{\omega_2}\ps \odagg \psi_{\omega_1} - \ps \psi_{\omega_1} \odagg \psi_{\omega_2})$. Equating the two terms, acting with the Teukolsky operator on the mode functions, and using BL coordinates, yields
\begin{eqnarray}
 \langle \! \langle \psi_{\omega_1},\psi_{\omega_2} \rangle \! \rangle &=&
 \frac{4 i M^{4/3}}{\omega_2 - \omega_1} \int \mathrm{d}\phi \mathrm{d}r \mathrm{d}\theta \sin\theta \Delta^{-2} 
 \nonumber \\ 
 && \times
 \left(\psi_{\omega_2} \hat T_{\omega_1} \psi_{\omega_1} - \psi_{\omega_1}\hat T_{\omega_2} \psi_{\omega_2}  \right).
\end{eqnarray}
Using the self-adjointness property of the Sturm-Liouville weighted integral, we can rewrite the term in brackets as $\psi_{\omega_1} (\hat T_{\omega_1} - \hat T_{\omega_2}) \psi_{\omega_2}$. Using \eqref{reflection in BL coordinates} and restoring the covariant integral $\surf t^a$, we obtain
\begin{equation}\label{eq:SL-off-diagonal}
  \langle \! \langle \psi_{\omega_1}, \psi_{\omega_2} \rangle \! \rangle = -i\left\langle \psi_{\omega_1}, \frac{\hodagg_{\omega_2}- \hodagg_{\omega_1}}{\omega_2 - \omega_1}\psi_{\omega_2}\right\rangle.
\end{equation}
On mode solutions, this equivalence was first obtained in \cite{Ma:2024qcv}. Notice that in the limit $\omega_2\to\omega_1$, we recover \eqref{eq: norm as operator derivative}.

We see from \eqref{eq:SL-off-diagonal} that orthogonality between QNMs with different frequencies can be expressed as the off-diagonal matrix element of the weight $i(\hodagg_{\omegan}- \hodagg_{\omeganp})/(\omegan - \omeganp)$ with respect to the SL product. The SL product itself is not orthogonal on QNMs. More generally, we will see in Sec.~\ref{sec:Modeshifts} that projection coefficients (defined with respect to the bilinear form) are naturally expressed in terms of matrix elements using the SL product. In quantum mechanics, no such distinction occurs, since for $T(E) \equiv H - E$ the weight is trivial, $\partial_E T = -\openone$; here the distinction is a consequence of the inherently nonlinear eigenvalue problem noted in Sec.~\ref{sec:gf}.

\section{Frequency shifts from the QNM bilinear form}\label{sec: Formalism}

We now use the bilinear form to develop a Teukolsky-based framework analogous to Schr\"odinger perturbation theory, suited to compute shifts in the QNM spectrum.

\subsection{Modified Teukolsky equations}

Spectral shifts can arise due to several factors:
(i) the equations for the gravitational perturbations differ from those of GR as the underlying theory is a  modified theory of gravity continuously connected to GR via some small coupling parameter $\zeta$ \cite{Blazquez-Salcedo:2024dur,Cano:2024ezp,Chung:2023wkd,Chung:2024ira,Chung:2024vaf,Chung:2025gyg,Li:2023ulk,Li:2025fci,Weller:2024qvo,Aly:2026otj}; (ii) the spacetime is a GR BH not of the Kerr type, like Kerr-AdS or Kerr-Newman whose spectrum depends on an extra parameter such as a cosmological constant or an electric charge \cite{Zimmerman:2014aha,Arnaudo:2025uos}; (iii) time-independent modifications to the Kerr metric due to the presence of an environment, whose density offers a natural expansion parameter $\zeta$ \cite{Pezzella:2024tkf,DellaRocca:2025xwz,Spieksma:2024voy}.
All these departures from vacuum GR, captured by a single characteristic dimensionless parameter $\zeta$, affect the ringdown through modifications of the Teukolsky equation \cite{Li:2022pcy,Cano:2023tmv,Dyson:2025dlj,Li:2025ffh,Dyson:2026ddd}. 
Thus, to leading order, we write the %
perturbation to the Weyl scalar as
\begin{equation}
\psi = \psi^{(0)}+\zeta \psi^{(1)} + O(\zeta^2) .
\end{equation}
Note that we highlight the order in the beyond-vacuum-GR parameter $\zeta$, while working to linear order in the field perturbation. %

Without loss of generality, we consider the spin $s=-2$ case. Suppose first for simplicity that for all $\zeta$, a Teukolsky equation holds,
\begin{equation}
    \odagg(\zeta) \psi(\zeta) = 0.
\end{equation}
This corresponds, for instance, to linearized GR off of the Kerr-(anti-)de Sitter family, itself parametrized by $\zeta$. (We will generalize this below.) To leading order in $\zeta$, we have the Teukolsky equation about the $\zeta=0$ background,
\begin{equation}
  \label{eq: Teukolsky equation for Kerr spacetime}
  \odaggzero \psi^{(0)} = 0.
\end{equation}
Expanding in $\zeta$ and equating order by order yields
\begin{eqnarray}
\label{eq: first order modified teukolsky equation type D}
    \odaggzero \psi^{(1)} &=& -\odaggone\psi^{(0)} ,\\
\label{eq: second order modified teukolsky equation type D}
    \odaggzero \psi^{(2)} &=& -\odagg_{(2)}\psi^{(0)} - 2\odaggone\psi^{(1)} .
\end{eqnarray}
We see that perturbative corrections satisfy an inhomogeneous equation, where the left hand side involves the background Teukolsky operator, and the source on the right hand side is made up of corrections to the operator acting on lower-order fields.

More generally, the perturbations $\psi^{(1)}$ and $\psi^{(2)}$ satisfy modified Teukolsky equations, which involve more complicated source terms, but maintain the same structure as above. %
Modified Teukolsky equations have been shown to encompass a wide range of bGR theories continuously connected to GR by a single coupling constant and with up to a single additional scalar degree of freedom~\cite{Li:2022pcy,Cano:2023tmv}. Ref.~\cite{Li:2023ulk} showed how, for the purpose of computing spectral shifts to the Kerr spectrum, the relevant equation can be cast in the schematic form 
\begin{equation}
\label{eq: first order modified teukolsky equation}
    \odaggzero \psi_{\boldsymbol n}^{(1)} = -\mathcal{S}^{(1)}[\psi_{\boldsymbol n}^{(0)} ].
\end{equation}
Here $\mathcal{S}^{(1)}$ is an integro-differential operator acting on $\psi^{(0)}$. %
In general, the source depends on the metric perturbation---not the Weyl scalar---so it is necessary to reconstruct the metric perturbation from $\psi^{(0)}$. Further, since the metric is real, $\mathcal{S}^{(1)}$ involves complex conjugation; acting on the mode $\psi_{lmnp}^{(0)}$ gives a mode $\psi_{l-mnp}^{(0)}$ with frequency $-\omega_{lmnp}^{*(0)}$. Due to the specific structure of the metric reconstruction operators, both $\psi_{lmnp}^{(0)}$ and $\psi_{l-mn}^{(0)}$ need to be considered as they produce resonances that are then interpreted as frequency shifts of the same $\omega^{(0)}_{lmnp}$. 
The problem is thus similar to degenerate perturbation theory, as worked out in \cite{Li:2023ulk}. In the remainder of this work, for simplicity, we set aside this further complication and assume that $\mathcal{S}^{(1)}$ is a linear operator acting on $\psinzero$ alone. We expect our formalism to be readily generalizable to the degenerate case by following \cite{Li:2023ulk}.

The effect of the back-reaction of the gravitationally-driven piece of the extra dynamical field
on $\psin^{(1)}$ is also taken into account within $\mathcal{S}^{(1)}$. For more details on the construction of $\mathcal{S}^{(1)}$, see \cite{Li:2023ulk}. We expect a similar structure to hold even beyond the assumptions of Ref.~\cite{Li:2022pcy}, e.g.~when the additional degrees of freedom are not scalar.

At higher order, we expect the same structure to hold, yielding %
\begin{equation}
  \label{eq: second order modified teukolsky equation}
  \odaggzero \psin^{(2)} = -\mathcal{S}^{(2)}\psin^{(0)} - 2\mathcal{S}^{(1)}\psin^{(1)},
\end{equation}
where we have again assumed that $\mathcal{S}^{(2)}$ is a linear operator.
However, for second order and beyond, it is necessary to also perform sourced metric reconstruction to obtain $\psi^{(1)}$, e.g., the GHZ metric reconstruction scheme \cite{Green:2019nam}, rather than the usual CCK scheme \cite{Chrzanowski:1975wv, Kegeles:1979an}.  %
A generic $k$-th order modified Teukolsky equation is then expected to have the form
\begin{equation}
  \label{eq: general k-th order MTE}
  \odagg_{(0)}\psin^{(k)}=-\sum_{j=0}^{k-1}\frac{k!}{j!(k-j)!}\mathcal{S}_{(k-j)}\psin^{(j)}.
\end{equation}

\subsection{Frequency shifts to arbitrary order}\label{subsec: Shifts}

Consider the balance law \eqref{Master Equation}, with $\psi_1 = \psinzero$ a Kerr QNM and $\psi_2 = \psin$ a QNM of the perturbed spacetime, continuously connected to $\psinzero$ as $\zeta \rightarrow 0$,
\begin{equation}\label{eq:conservation-law-qnm}
  \frac{\mathrm{d}}{\mathrm{d}t} \langle\! \langle \psinzero , \psin  \rangle\! \rangle =  \int_{\Sigma_t} \mathrm{d}\Sigma_a t^a \nabla_b \pi^b [\ps \psinzero , \psin].
\end{equation}
We perform a power series expansion of $\psin$ and its associated frequency $\omegan$ about the respective background quantities,
\begin{eqnarray}
  \psin &=& \psin^{(0)} + \zeta \psinone + \frac{\zeta^2}{2} \psin^{(2)} + \ldots , \\
  \omegan &=& \omegan^{(0)} + \zeta \omegan^{(1)} + \frac{\zeta^2}{2} \omegan^{(2)} + \ldots .
\end{eqnarray}
For the left-hand side of \eqref{eq:conservation-law-qnm}, we have
\begin{eqnarray}
  \label{LHS}
  \frac{\mathrm{d}}{\mathrm{d}t} \langle\! \langle \psinzero , \psin  \rangle \!\rangle &=& -i(\omegan - \omegan^{(0)}) \langle\! \langle \psinzero , \psin  \rangle\! \rangle  \nonumber \\
  &=& -i \zeta \omegan^{(1)} \normQNM + O(\zeta^2).
\end{eqnarray}
For the right-hand side, we use the definition of the symplectic current \eqref{symplectic current} to express the integrand in terms of the Teukolsky operator,
\begin{eqnarray}
  \label{RHS}
  &&  \nabla_b \pi^b [\ps \psinzero, \psin] \nonumber \\
  &=&  \mathcal{O}_{(0)}(\ps \psinzero) \psin - (\ps \psinzero ) \odaggzero \psin.
\end{eqnarray}
The first term vanishes because $\psinzero$ solves the adjoint Teukolsky equation, hence $\ps \psinzero$ solves the Teukolsky equation. In the second term,
\begin{equation}
  \odaggzero \psin = \zeta \odaggzero \psin^{(1)} + O(\zeta^2)  = - \zeta \mathcal{S}_{(1)} \psi_{\boldsymbol n}^{(0)} + O(\zeta^2),
\end{equation}
where we used the modified Teukolsky equation \eqref{eq: first order modified teukolsky equation} in the last step.

Equating the left and right hand sides, we obtain the expression for the first-order frequency shift,
\begin{equation}
  \label{first order frequency shift}
  \omega_{\boldsymbol n}^{(1)} = i \frac{\int_{\Sigma_t} \mathrm{d}\Sigma_a t^a (\ps \psi_{\boldsymbol n}^{(0)}) \mathcal{S}_{(1)}\psi^{(0)}_{\boldsymbol n} }{\langle \!\langle \psi_{\boldsymbol n}^{(0)} ,  \psi_{\boldsymbol n}^{(0)}  \rangle\! \rangle} \, .
\end{equation}
This formula coincides with the one obtained via the \textit{eigenvalue perturbation method} \cite{Zimmerman:2014aha,Mark:2014aja,Hussain:2022ins,Li:2023ulk} (see e.g.,~Eq.~(86) of \cite{Ma:2024qcv}), which is based on the SL product rather than the bilinear form. Indeed, substituting the norm identity \eqref{eq: norm as operator derivative} into the denominator of \eqref{first order frequency shift}, we obtain
\begin{equation}
  \omegan^{(1)} = - \frac{\langle \psinzero, \mathcal{S}_{(1)}\psinzero \rangle}{\langle \psinzero, \partial_\omega \hodagg_{(0)}\big|_{\omegan^{(0)}}  \psi_{\boldsymbol n}^{(0)}\rangle},
\end{equation}
a ratio of SL-product matrix elements of the source and of $\partial_\omega \hodagg_{(0)}\big|_{\omegan^{(0)}}$. This is precisely the structure of the eigenvalue-perturbation-theory formula, which is recovered explicitly in BL coordinates and the Kinnersley frame.

If we also keep terms of order $\zeta^2$, we get an expression for the second order frequency shift,
\begin{eqnarray}
\label{second order shift}
  \omegan^{(2)} &=& i \frac{\surf t^a \ps \psinzero \left( \mathcal{S}_{(2)} \psinzero + 2 \mathcal{S}_{(1)} \psinone  \right)}{\normQNM} \nonumber \\
  && -2 \omegan^{(1)} \frac{\projnone}{\normQNM}.
\end{eqnarray} %
This equation, although simple and similar in structure to the formula for the first order frequency shift, depends on the first-order mode shift $\psinone$, whose computation is discussed in the next section. Notice that the second-order frequency shift is invariant under $\psinone \rightarrow \psinone + \beta \psinzero$, for constant $\beta$. Indeed, under this transformation we have
\begin{equation}
\label{invariance of second order frequency shift}
  \omegan^{(2)} \rightarrow \omegan^{(2)} - 2 \beta \omegan^{(1)} +2 \beta i \frac{\surf \ps \psinzero \mathcal{S}_{(1)} \psinzero }{\normQNM} .
\end{equation}
Recognizing that the last term is $\omegan^{(1)}$ from Eq.~\eqref{first order frequency shift}, we prove the invariance. 

We also note that the projection $\projnone$ appearing in \eqref{second order shift} is \emph{not} time independent, since $\psinone \notin \ker \odaggzero$. Its time dependence cancels against that of the integral featuring $\mathcal{S}_{(1)}\psinone$, leaving $\omegan^{(2)}$ constant, as we show explicitly in App.~\ref{app: second order details}. The origin of this time dependence (a secular term in the mode shift) is discussed in Sec.~\ref{sec:Modeshifts}. %

Assuming the existence of the $k$-th order modified Teukolsky equation \eqref{eq: general k-th order MTE}, one can generalize the calculation to any order.
By a similar calculation as above, keeping terms to $O(\zeta^k)$, we obtain a formula for the $k$-th order frequency shift in terms of the unperturbed mode $\psinzero$, and the mode shifts $\psinone,\psin^{(2)},...,\psin^{(k-1)}$,
\begin{eqnarray}
\label{eq: generic k-th order frequency shift}
  \omegan^{(k)} &=& i \frac{ \surf t^a \ps \psinzero \sum_{j=0}^{k-1}\frac{k!}{j!(k-j)!}\mathcal{S}_{(k-j)}\psin^{(j)}}{\normQNM} \nonumber \\
  && -\sum_{j=1}^{k-1}\frac{k!}{j!(k-j)!}\omegan^{(j)}\frac{\langle \! \langle \psinzero, \psin^{(k-j)} \rangle \! \rangle}{\normQNM}.
\end{eqnarray}
At this point, quantum mechanical perturbation theory would proceed by expanding
the mode shift $\psinone$ in the complete set of unperturbed eigenstates, thereby
reducing \eqref{second order shift} to a closed-form spectral sum and integral over matrix
elements of the perturbation. It is precisely this step that fails for QNMs, as
we describe in the next section.

\section{Mode shifts and the breakdown of the QNM spectral expansion}\label{sec:Modeshifts}

To complete the perturbative framework, it is necessary to compute shifts to the mode functions, which enter the frequency-shift formulas at second and higher order. As we will see, relative to the standard quantum-mechanical treatment, the computation differs in two ways. First, because the eigenvalue problem is nonlinear in $\omega$, the frequency shift leaks into the mode shift itself, as a secular growth in $t$, and into the expansion coefficients, through the projection of the secular term on background QNMs.
Second, since there is no spectral theorem for QNMs, the decomposition of the mode shift over QNMs plus a continuum piece is only formal, and the QNM sum in fact diverges (see example in Sec.~\ref{sec: PT potential}).

Our expansion mirrors the standard quantum-mechanical treatment, which we briefly recall. In quantum mechanics, for a small, time-independent perturbation of the Hamiltonian, one can write
\begin{eqnarray}
\label{eq: QM spectral decomposition of mode shift}
  \ket{\psinone} &=& \sum_{\boldsymbol n'\neq \boldsymbol n} \frac{\braket{\psinpzero|\psinone}}{\braket{\psinpzero|\psinpzero}} \ket{\psinpzero} \nonumber \\
  && + \sum_{a}\int \mathrm{d}E \braket{\psi_a(E)|\psinone}\ket{\psi_a(E)},
\end{eqnarray}
The first term expresses the projection of the perturbed state $\ket{\psinone}$ onto bound states $\ket{\psinpzero}$. This involves the projection coefficients, which are given in terms of the perturbation $V$ of the Hamiltonian as $\braket{\psinpzero|\psinone} = \braket{\psinpzero|V|\psinzero}/(E_n - E_{n'})$. The second term, involving the integral over the continuous energy spectrum, is associated to unbound states. (Here, $a$ denotes discrete quantum numbers, e.g., the angular numbers $l$, $m$.) Neglecting these unbound states will render the expansion incomplete. Likewise, as we will see here and in the examples to follow, for QNMs it is necessary to include this continuum contribution. Note that the discrete sum omits $\boldsymbol{n}' = \boldsymbol{n}$; this is proportional to $\ket{\psinzero}$ and hence can always be absorbed into normalization of the background state.

It is most straightforward to derive the spectral decomposition for the mode shift using Green's function techniques, as these naturally encompass the continuum contributions. We do so in the following subsection. Afterwards, we will write the QNM component of the spectral decomposition in terms of the bilinear form, and make explicit the analogies with (and the departures from) quantum mechanics.

\subsection{Mode shifts from the Green's function}

\begin{figure*}[t]
\centering
\subfloat
{
  \tikzset{
  midarrow/.style={
    postaction={decorate},
    decoration={markings, mark=at position 0.6 with {\arrow{Stealth}}}
  }
}

\tikzset{
  midarrowcircle/.style={
    postaction={decorate},
    decoration={markings, mark=at position 0.7 with {\arrow{Stealth}}}
  }
}

\begin{tikzpicture}[>=Stealth, scale=1.1]

\def\R{3.2}        %
\def\eps{0.18}     %
\def\a{1.15}       %
\def\b{1.10}       %
\def\c{2.05}       %
\def\d{0.50}       %
\def\e{0.65}       %
\def\f{2.10}       %
\def\rp{0.22}      %
\def\Ycut{3.4}     %
\def\dth{10}       %

\tikzset{contour/.style={rossos,
thick, line cap=round, line join=round}}

\tikzset{
  midarrow/.style={
    postaction={decorate},
    decoration={markings, mark=at position 0.55 with {\arrow{Stealth[length=2.1mm,width=2.1mm]}}}
  }
}

\draw[->] (-3.4,0) -- (3.2,0) node[right] {$\Re \omega$};
\draw[->] (0,-3.4) -- (0,3.2) node[above] {$\Im \omega$};

\draw[very thick, decorate,
      decoration={zigzag, segment length=6pt, amplitude=1.2pt}]
  (0,-0.06) -- (0,-\Ycut);

  \coordinate (rH) at (0,0);
  \fill (rH) circle (1.2pt) ;

\coordinate (pR) at (\a,-\b);
\coordinate (pL) at (-\a,-\b);
\coordinate (pR1) at (\c,-\d);
\coordinate (pL1) at (-\c,-\d);
\coordinate (pR2) at (\e,-\f);
\coordinate (pL2) at (-\e,-\f);

\fill (pR) circle (1.3pt) node[below right] {$\omeganp^{(0)}$};
\fill (pL) circle (1.3pt) node[below left]  {$\omegan^{(0)}$};
\fill (pR1) circle (1.3pt) node[below right] {};
\fill (pL1) circle (1.3pt) node[below left]  {};
\fill (pR2) circle (1.3pt) node[below right] {};
\fill (pL2) circle (1.3pt) node[below left]  {};

\pgfmathsetmacro{\delta}{asin(\eps/\R)}               %
\pgfmathsetmacro{\angA}{-90+\delta}                  %

\draw[contour] (pL) ++(\angA:\rp) arc (\angA:{90-\dth}:\rp);
\draw[contour, -{Stealth[length=2.0mm,width=2.0mm]}]
  (pL) ++({90-\dth}:\rp) arc ({90-\dth}:{90+\dth}:\rp);
\draw[contour] (pL) ++({90+\dth}:\rp) arc ({90+\dth}:{360}:\rp);

\end{tikzpicture}
}
\hfill
\subfloat
{
  \tikzset{
  midarrow/.style={
    postaction={decorate},
    decoration={markings, mark=at position 0.6 with {\arrow{Stealth}}}
  }
}

\tikzset{
  midarrowcircle/.style={
    postaction={decorate},
    decoration={markings, mark=at position 0.7 with {\arrow{Stealth}}}
  }
}

\begin{tikzpicture}[>=Stealth, scale=1.1]

\def\R{3.2}        %
\def\eps{0.18}     %
\def\a{1.15}       %
\def\b{1.10}       %
\def\c{2.05}       %
\def\d{0.50}       %
\def\e{0.65}       %
\def\f{2.10}       %
\def\rp{0.22}      %
\def\Ycut{3.4}     %
\def\dth{10}       %

\tikzset{contour/.style={rossos,
thick, line cap=round, line join=round}}

\tikzset{
  midarrow/.style={
    postaction={decorate},
    decoration={markings, mark=at position 0.55 with {\arrow{Stealth[length=2.1mm,width=2.1mm]}}}
  }
}

\draw[->] (-3.4,0) -- (3.2,0) node[right] {$\Re \omega$};
\draw[->] (0,-3.4) -- (0,3.2) node[above] {$\Im \omega$};

\draw[very thick, decorate,
      decoration={zigzag, segment length=6pt, amplitude=1.2pt}]
  (0,-0.06) -- (0,-\Ycut);

  \coordinate (rH) at (0,0);
  \fill (rH) circle (1.2pt) ;

\coordinate (pR) at (\a,-\b);
\coordinate (pL) at (-\a,-\b);
\coordinate (pR1) at (\c,-\d);
\coordinate (pL1) at (-\c,-\d);
\coordinate (pR2) at (\e,-\f);
\coordinate (pL2) at (-\e,-\f);

\fill (pR) circle (1.3pt) node[below right] {$\omeganp^{(0)}$};
\fill (pL) circle (1.3pt) node[below left]  {$\omegan^{(0)}$};
\fill (pR1) circle (1.3pt) node[below right] {};
\fill (pL1) circle (1.3pt) node[below left]  {};
\fill (pR2) circle (1.3pt) node[below right] {};
\fill (pL2) circle (1.3pt) node[below left]  {};

\pgfmathsetmacro{\delta}{asin(\eps/\R)}               %
\pgfmathsetmacro{\angA}{-90+\delta}                  %
\pgfmathsetmacro{\angB}{270-\delta}                  %
\pgfmathsetmacro{\ybot}{-sqrt(\R*\R - \eps*\eps)}    %

\draw[contour, midarrow] (\eps,0) -- (\eps,\ybot);

\draw[contour, midarrow] (-\eps,\ybot) -- (-\eps,0);

\draw[contour] (0,0) ++(180:\eps) arc (180:{90+\dth}:\eps);
\draw[contour]
  (0,0) ++({90+\dth}:\eps) arc ({90+\dth}:{90-\dth}:\eps);
\draw[contour] (0,0) ++({90-\dth}:\eps) arc ({90-\dth}:0:\eps);

\draw[contour] (0,0) ++(\angA:\R) arc (\angA:{90-\dth}:\R);
\draw[contour, -{Stealth[length=2.1mm,width=2.1mm]}]
  (0,0) ++({90-\dth}:\R) arc ({90-\dth}:{90+\dth}:\R);
\draw[contour] (0,0) ++({90+\dth}:\R) arc ({90+\dth}:\angB:\R);

\node at (2.15,2.65) {$\gamma$};

\draw[contour] (pR) ++(\angA:\rp) arc (\angA:{90-\dth}:\rp);
\draw[contour, -{Stealth[length=2.0mm,width=2.0mm]}]
  (pR) ++({90+\dth}:\rp) arc ({90+\dth}:{90-\dth}:\rp);
\draw[contour] (pR) ++({90+\dth}:\rp) arc ({90+\dth}:{360}:\rp);

\draw[contour] (pR1) ++(\angA:\rp) arc (\angA:{90-\dth}:\rp);
\draw[contour, -{Stealth[length=2.0mm,width=2.0mm]}]
  (pR1) ++({90+\dth}:\rp) arc ({90+\dth}:{90-\dth}:\rp);
\draw[contour] (pR1) ++({90+\dth}:\rp) arc ({90+\dth}:{360}:\rp);

\draw[contour] (pR2) ++(\angA:\rp) arc (\angA:{90-\dth}:\rp);
\draw[contour, -{Stealth[length=2.0mm,width=2.0mm]}]
  (pR2) ++({90+\dth}:\rp) arc ({90+\dth}:{90-\dth}:\rp);
\draw[contour] (pR2) ++({90+\dth}:\rp) arc ({90+\dth}:{360}:\rp);

\draw[contour] (pL1) ++(\angA:\rp) arc (\angA:{90-\dth}:\rp);
\draw[contour, -{Stealth[length=2.0mm,width=2.0mm]}]
  (pL1) ++({90+\dth}:\rp) arc ({90+\dth}:{90-\dth}:\rp);
\draw[contour] (pL1) ++({90+\dth}:\rp) arc ({90+\dth}:{360}:\rp);

\draw[contour] (pL2) ++(\angA:\rp) arc (\angA:{90-\dth}:\rp);
\draw[contour, -{Stealth[length=2.0mm,width=2.0mm]}]
  (pL2) ++({90+\dth}:\rp) arc ({90+\dth}:{90-\dth}:\rp);
\draw[contour] (pL2) ++({90+\dth}:\rp) arc ({90+\dth}:{360}:\rp);

\end{tikzpicture}
}
\caption{Complex contours featured in the derivation of Eqs.~\eqref{eq: Cauchy identity} and \eqref{eq: frequency integral of green function - exploiting the contour}. %
The residue theorem allows to deform the complex path on the left to the contours on the right.  %
}
\label{fig: complex contour for mode shift calculation}
\end{figure*}
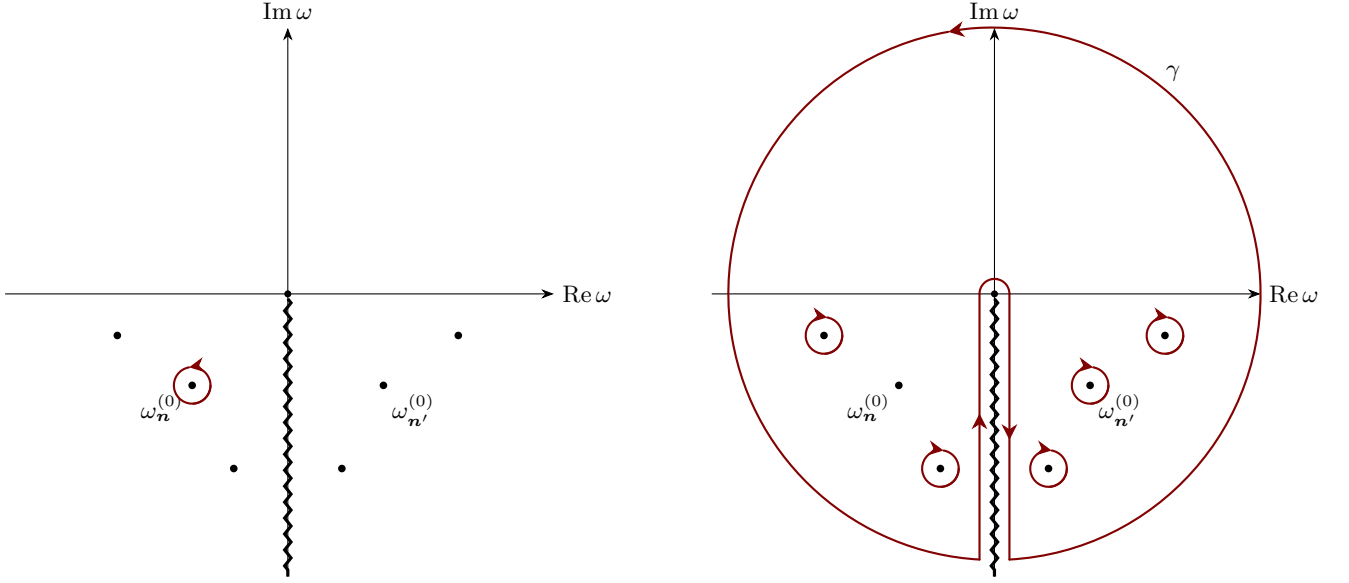

To derive the spectral decomposition of the mode shift $\psinone$, we start by considering the leading-order modified Teukolsky equation \eqref{eq: first order modified teukolsky equation}, written in terms of the separable Teukolsky operator $T = \Sigma \odaggzero$,
\begin{equation}
\label{eq: first order modified teukolsky equation - BL coordinates}
    T\psinone = - \Sigma \mathcal{S}^{(1)}\psinzero.
\end{equation}
We decompose the field as $\psin(\zeta) = e^{-i \omegan(\zeta)t}\chi(\boldsymbol{x},\zeta)$ and work in frequency domain. The first order field $\psinone$ then takes the form, 
\begin{equation}\label{eq:psinone-secular}
    \psinone(t,\boldsymbol{x}) = -i \omegan^{(1)} t \, \psinzero(t,\boldsymbol{x}) + e^{-i \omegan^{(0)}t} \chi^{(1)}_{\boldsymbol n}(\boldsymbol{x}).
\end{equation}
with $\boldsymbol{x}=(r,\theta,\phi)$. Notice that the expansion in $\zeta$ introduces a secular term that grows linearly in $t$.

Inserting the expansion \eqref{eq:psinone-secular} into \eqref{eq: first order modified teukolsky equation - BL coordinates} yields an equation for the spatial part $\chi^{(1)}_{\boldsymbol n}$,
\begin{eqnarray}
\label{eq: equation for chi1}
  \hat T_{\omegan^{(0)}} \chi^{(1)}_{\boldsymbol n} &=& - \left( \Sigma \hat{\mathcal{S}}^{(1)}_{\omegan^{(0)}} + \omegan^{(1)} \partial_{\omega}\hat T\big|_{\omegan^{(0)}}  \right)\chi^{(0)}_{\boldsymbol n} \nonumber \\
  &=& Q^{(1)}_{\boldsymbol n}\chi^{(0)}_{\boldsymbol n},
\end{eqnarray}
 where we used 
\begin{equation}\label{eq:identity_dT}
    T(-i \omegan^{(1)}t \psinzero) = \omegan^{(1)} e^{-i \omegan^{(0)}t}\partial_{\omega}\hat T\big|_{\omegan^{(0)}} \chi^{(0)}_{\boldsymbol n}.
\end{equation}
Note that the frequency-domain Teukolsky operator on the left hand side of \eqref{eq: equation for chi1} is evaluated at a QNM frequency. At first glance the equation looks not suited to be solved using the Green function of the operator, as the latter has simple poles at those frequencies. To avoid this, we
define the auxiliary function $\chi^{(1)}_{\omega}$, with $\omega$ a generic frequency, as the solution of 
\begin{equation}
\label{eq: auxiliary equation for chi1}
    \hat T_{\omega}\chi^{(1)}_{\omega} = Q^{(1)}_{\boldsymbol n} \chi^{(0)}_{\boldsymbol n}.
\end{equation}
Then, in App.~\ref{app: integral representation of mode shift}, we show that the solution $\chi^{(1)}_\omega$ is regular at $\omega = \omegan^{(0)}$ even when expressed in terms of the Green function. Using the residue theorem, we can express $\chi^{(1)}_{\boldsymbol n}$ as an integral around a circle centered on $\omegan^{(0)}$ as
\begin{eqnarray}
\label{eq: Cauchy identity}
  \chi^{(1)}_{\boldsymbol n} &=& \frac{1}{2 \pi i} \int_{\partial B_{\epsilon}(\omegan^{(0)})} \frac{\chi^{(1)}_{\omega}}{\omega - \omegan^{(0)}} \mathrm{d}\omega  \\
  &=& \frac{1}{2 \pi i} \int_V \mathrm{d}^3\boldsymbol{x}' Q^{(1)}_{\boldsymbol n}(\boldsymbol{x}')\chi^{(0)}_{\boldsymbol n}(\boldsymbol{x}') \int_{\partial B_{\epsilon}(\omegan^{(0)})} \frac{g(\omega,\boldsymbol{x},\boldsymbol{x}')}{\omega - \omegan^{(0)}} \mathrm{d}\omega, \nonumber
\end{eqnarray}
where the contour $\partial B_{\epsilon}(\omega_n^{(0)})$ runs anticlockwise around the pole (see left panel of Fig.~\ref{fig: complex contour for mode shift calculation}). 
On the second line we replaced $\chi^{(1)}_\omega$ with the solution
of \eqref{eq: auxiliary equation for chi1} (with zero initial data) in terms of the Green's function $g(\omega,\boldsymbol{x},\boldsymbol{x}')$ of the operator $\hat T_{\omega}$. We provide the Green's function in App.~\ref{app: integral representation of mode shift}, as well as the relevant measure $\mathrm{d}^3\boldsymbol{x}'$. 

To obtain the spectral decomposition of $\chi_{\boldsymbol n}^{(1)}$, we re-express the integral around the $\omegan^{(0)}$-pole as an integral around all of the other poles, as well as along the branch-cut and arc at infinity. %
When decomposed in $l',m'$ modes, the last line of \eqref{eq: Cauchy identity} contains $g_{l'm'}(\omega,r,r')$, see Eq.~\eqref{eq:GF_radial_main}, which has simple poles in the complex-$\omega$ plane at the quasinormal frequencies $\omega_{\boldsymbol{n}'}^{(0)}$ and a branch cut running along the negative imaginary axis. The residue theorem implies that the integral along the contour $\partial B_{\epsilon}(\omega_{\boldsymbol{n}}^{(0)})$ is equal to an integral along the contour $\gamma$ depicted in the right panel of Fig.~\ref{fig: complex contour for mode shift calculation} plus the sum of the residues associated with the poles at $\omega_{\boldsymbol{n}'}^{(0)}$ with $\boldsymbol{n}'\neq \boldsymbol{n}$.
Following Fig.~\ref{fig: complex contour for mode shift calculation}, the frequency integral can then be written as
\begin{equation}
\label{eq: frequency integral of green function - exploiting the contour}
  \left( \sum_{l'm'}\int_{\gamma} - \sum_{\boldsymbol n' \neq \boldsymbol n}\int_{\partial B_{\epsilon}(\omeganp^{(0)})}\right)\frac{g^\Omega_{l'm'}(\omega,\Omega,\Omega')g_{l'm'}(\omega,r,r')}{\omega - \omegan^{(0)}} \mathrm{d}\omega,
\end{equation}
where we used the decomposition of the Green function provided in App.~\ref{app: integral representation of mode shift}.
This corresponds to a split into continuous and QNM contributions, $\chi^{(1)}_{\boldsymbol n} =   \chi^{(1)}_{\gamma} +\chi^{(1)}_{\text{QNM}}$.

\begin{widetext}
Using the expression \eqref{eq:GF_radial_main} for the radial Green's function, along with the residue theorem, we obtain
\begin{eqnarray}
\label{eq: frequency integral of green function - applying residue theorem}
   \int_{\partial B_{\epsilon}(\omeganp^{(0)})}\frac{g^\Omega_{l'm'}(\omega,\Omega,\Omega')g_{l'm'}(\omega,r,r')}{\omega - \omegan^{(0)}} \mathrm{d}\omega 
  = \frac{8 \pi M^{4/3}\Delta'^{-2}}{\Delta \omega_{\boldsymbol{nn}'}^{(0)}} \frac{R_{\boldsymbol n'}^{(0)}(r)R_{\boldsymbol n'}^{(0)}(r')}{\normp}g^\Omega_{l'm'}(\omeganp^{(0)},\Omega,\Omega').
\end{eqnarray}
The QNM component of $\chi_{\boldsymbol n}^{(1)}$ is therefore
\begin{eqnarray}\label{eq:qnm-modeshift-1}
  \chi^{(1)}_{\text{QNM}} &=& \sum_{\mathbf{n'}\neq \boldsymbol n}\frac{4M^{4/3}i}{\Delta\omega_{\boldsymbol{nn}'}^{(0)}} \frac{e^{i m' \phi}S^{(0)}_{\boldsymbol n'}(\theta)R^{(0)}_{\boldsymbol n'}(r)}{\normp} \nonumber 
 \int \mathrm{d}r' \mathrm{d}\theta' \mathrm{d}\phi' \sin\theta' \Delta'^{-2}e^{-i m' \phi'}S^{(0)}_{\boldsymbol n'}(\theta')R^{(0)}_{\boldsymbol n'}(r') \nonumber \\
   && \times
   \left( \Sigma' \hat{\mathcal{S}}^{(1)}_{\omegan^{(0)}}+\omegan^{(1)}\partial_{\omega}\hat T\big|_{\omegan^{(0)}} \right)e^{i m \phi'}S^{(0)}_{\boldsymbol n}(\theta')R^{(0)}_{\boldsymbol n}(r'),
\end{eqnarray}
where $\Delta\omega_{\boldsymbol{nn}'}^{(0)} \equiv \omegan^{(0)} - \omeganp^{(0)}$, and where we used the angular Green's function \eqref{eq: angular GF}.
It is useful to reintroduce the spacetime mode functions in the expansion above, in order to make contact with the bilinear form in the following subsection, 
\begin{eqnarray}
    \label{eq: spatial mode shift in terms of QNM sum}
  \chi^{(1)}_{\text{QNM}}e^{-i \omegan^{(0)}t}&=& \sum_{\boldsymbol n' \neq \boldsymbol n} \frac{i}{\Delta\omega^{(0)}_{\boldsymbol{nn}'}}\frac{\psinpzero}{\normp}
    \left[ \surf t^a \ps \psinpzero \mathcal{S}^{(1)}\psinzero \right.
   + \left. \surf t^a \ps \psinpzero \odaggzero(-i \omegan^{(1)}t \psinzero) \right]
   \nonumber \\ 
   &=& \sum_{\boldsymbol n' \neq \boldsymbol n} \frac{i}{\Delta\omega^{(0)}_{\boldsymbol{nn}'}}\frac{\psinpzero}{\normp}
    \left[ \left\langle\psinpzero,S^{(1)}\psinzero\right\rangle +  \left\langle \psinpzero, \odaggzero (-i \omegan^{(1)}t \psinzero) \right\rangle \right].
\end{eqnarray}
Eq.~\eqref{eq: spatial mode shift in terms of QNM sum} is the QNM contribution to the spectral decomposition of the mode shift. 
We note that the sum also runs over all possible values of $m'$ and is in general not expected to collapse to $m'=m$. Indeed, a generic perturbation of the Kerr spacetime could break axisymmetry, mixing different $m$-modes.

For completeness, we also provide the expression of the continuous contribution $\chi^{(1)}_{ \gamma}$,
\begin{eqnarray}
\label{eq: continuous spectral part of modeshift}
  \chi^{(1)}_{ \gamma}= -\frac{1}{2 \pi i} \sum_{l'm'} \int_{\gamma}\frac{\mathrm{d}\omega}{\omega-\omegan^{(0)}}\int \mathrm{d}r' \mathrm{d}\theta' \mathrm{d}\phi' \sin\theta' \Delta'^{-2} 
   \cdot Q_{\boldsymbol n}^{(1)}(\boldsymbol{x}')\chi^{(0)}_{\boldsymbol n}(\boldsymbol{x}')g^\Omega_{l'm'}(\omega,\Omega,\Omega')g_{l'm'}(\omega,r,r').
\end{eqnarray}
\end{widetext}

\subsection{Mode shifts from the bilinear form}\label{sec:Modeshifts-bilinear}

Consider the spectral decomposition \eqref{eq: spatial mode shift in terms of QNM sum}. The first integral appearing has the quantum-mechanical structure $\langle\psinpzero,S^{(1)}\psinzero\rangle/(\omegan^{(0)} - \omeganp^{(0)})$, with the matrix element taken in the SL product. It is natural to ask whether this is equal to a projection coefficient, as it would be in quantum mechanics. The relevant product for projection coefficients is the bilinear form, i.e., $\projnpone$ with $\boldsymbol{n}' \neq \boldsymbol{n}$, since the bilinear form gives QNM orthogonality. As noted at the end of Sec.~\ref{subsec: excitation coefficients}, in quantum mechanics the two products are equivalent, with the same product used for projection and for matrix elements. 

The computation of the projection coefficient is similar to the derivation of the first order frequency shift formula \eqref{first order frequency shift}, as it exploits the balance law \eqref{Master Equation}. 
Consider Eq.~\eqref{Master Equation} with $\psi_1 = \psinpzero$ and $\psi_2 = \psin$,
     \begin{equation}
         \frac{\mathrm{d}}{\mathrm{d}t} \langle\! \langle \psinpzero , \psin  \rangle\! \rangle =  \int_{\Sigma_t} \mathrm{d}\Sigma_a t^a \nabla_b \pi^b [\ps \psinpzero , \psin].
     \end{equation}
For the left-hand side we have, neglecting terms of order $\zeta^2$ and higher,
\begin{eqnarray}
\label{LHS 2}
  \frac{\mathrm{d}}{\mathrm{d}t} \langle\! \langle \psinpzero , \psin  \rangle\! \rangle &=& -i(\omegan - \omeganp^{(0)}) \langle\! \langle \psinpzero , \psin  \rangle\! \rangle \nonumber \\
  &=& -i \zeta (\omegan^{(0)}-\omeganp^{(0)}) \projnpone,
\end{eqnarray}
where in the last step we used the orthogonality of Kerr QNMs.
For the right-hand side, everything follows as in the proof of \eqref{first order frequency shift}. We find %
\begin{eqnarray} 
\label{projection coefficient}
    \langle \!\langle \psi_{\boldsymbol n'}^{(0)} ,  \psi_{\boldsymbol n}^{(1)}  \rangle\! \rangle &=& 
    \frac{i}{\Delta \omega_{\boldsymbol{nn}'}^{(0)}} \int_{\Sigma_t} \mathrm{d}\Sigma_a t^a (\ps \psi_{\boldsymbol n'}^{(0)}) \mathcal{S}_{(1)}\psi^{(0)}_{\boldsymbol n}
    \nonumber \\
    & = & \frac{i}{\Delta \omega_{\boldsymbol{nn}'}^{(0)}} \langle\psinpzero,S^{(1)}\psinzero\rangle,
\end{eqnarray}
which is the first matrix element that appears in \eqref{eq: spatial mode shift in terms of QNM sum}, similarly to Schr\"odinger perturbation theory.

Consider now the ${\boldsymbol n' = \boldsymbol n}$ projection coefficient. 
We can compute the time derivative of $\projnone$ using \eqref{Master Equation}, yielding
\begin{eqnarray}
  \frac{\d }{\d t} \projnone &=& - \surf t^a \ps \psinzero \odaggzero \psinone \nonumber \\
  &=& -i \omegan^{(1)} \normQNM,
\end{eqnarray}
where in the last step we have used Eq.~\eqref{eq: first order modified teukolsky equation} and the formula for the first order frequency shift \eqref{first order frequency shift}.
Integrating with respect to $t$ and using the fact that the right-hand side is time independent, we have
\begin{equation}
\label{eq: mode shift onto same mode}
   \frac{\projnone}{\normQNM}\psinzero  = -i \omegan^{(1)}t \psinzero + \alpha \psinzero, \quad \alpha \in \mathbb{C} \, .
\end{equation}

The second integral in \eqref{eq: spatial mode shift in terms of QNM sum} looks like the projection of the secular term onto a mode $\psinpzero$. This object is absent in QM and marks the first departure from the Schrödinger perturbation theory formula \eqref{eq: QM spectral decomposition of mode shift}. Since $\psinzero$ satisfies the adjoint Teukolsky equation, we can write
\begin{eqnarray}
\label{eq: identity with projection of secular term}
  &&\langle \psinpzero, \odaggzero (-i \omegan^{(1)}t \psinzero) \rangle
  \nonumber \\ &&\quad = 
  - \Delta \omega_{\boldsymbol{nn}'}^{(0)} \omegan^{(1)} \surf t^a \ps \psinpzero \frac{\Lambda}{\Sigma \Delta} \psinzero \nonumber \\
  &&\quad = i \Delta \omega_{\boldsymbol{nn}'}^{(0)} \langle \! \langle \psi_{\boldsymbol n'}^{(0)}, -i \omegan^{(1)} t \psinzero \rangle \! \rangle .
\end{eqnarray}
The last equality follows from the orthogonality of background QNMs.

With the projection coefficient \eqref{projection coefficient} at hand as well as the projection of the secular term \eqref{eq: identity with projection of secular term}, we can rewrite the spectral decomposition \eqref{eq: spatial mode shift in terms of QNM sum} as
\begin{widetext}
\begin{equation}
\label{eq: partial implicit spectral decomposition}
    \chi_{\text{QNM}}^{(1)}e^{-i \omegan^{(0)}t} =  \sum_{\boldsymbol n' \neq \boldsymbol n} \frac{\projnpone}{\normp} \psinpzero 
    + \sum_{\boldsymbol n' \neq \boldsymbol n} \frac{\langle \! \langle \psinpzero, -i \omegan^{(1)}t \psinzero \rangle \! \rangle}{\normp} \psinpzero. 
\end{equation}
Equivalently, adding to both sides the secular term and the continuous piece to restore $\psin^{(1)}$, and using \eqref{eq: mode shift onto same mode}, we have 
    \begin{equation}
    \label{eq: implicit spectral decomposition}
    \psinone = \psi^{(1)}_{\boldsymbol{n},\gamma} + \sum_{\boldsymbol n'} \frac{\projnpone}{\normp} \psinpzero 
    + \sum_{\boldsymbol n' \neq \boldsymbol n} \frac{\langle \! \langle \psinpzero, -i \omegan^{(1)}t \psinzero \rangle \! \rangle}{\normp} \psinpzero, 
\end{equation}
\end{widetext}
where we have defined $\psi^{(1)}_{\boldsymbol{n},\gamma} = e^{-i \omegan^{(0)}t}\chi^{(1)}_{\gamma}$.

Eq.~\eqref{eq: implicit spectral decomposition} is the spectral decomposition of the mode shift in terms of projections defined via the bilinear form. The first sum is analogous in form to the sum on bound states in the Schrödinger formula. However, in the QNM setting, it is not sufficient, and needs to be augmented with the projections of the secular term. Moreover, in quantum mechanics, one can always reabsorb the $\boldsymbol{n}' = \boldsymbol{n}$ term featured in the first sum into the background mode $\psinzero$. Here, the $\boldsymbol{n}' = \boldsymbol{n}$ term cannot be reabsorbed, as it is time-dependent and reproduces exactly the secular term \eqref{eq: mode shift onto same mode}.

In this section, we provided a formal spectral decomposition for the mode shift $\psinone$. However, we have not yet determined whether the featured QNM sums and the integral defining $\psi_{\boldsymbol n,\gamma}^{(1)}$ \eqref{eq: continuous spectral part of modeshift} are finite.
In section \ref{sec: PT potential}, in the context of Regge-Wheeler-like equations with a P{\"o}schl-Teller potential, we show numerically that the QNM sum actually diverges, marking an important departure from the familiar quantum mechanical perturbation theory. We discuss the spectral decomposition further and possible future developments in Sec.~\ref{sec:conclusions}.

As we expect this divergence to be a general feature, the practical way of computing %
$\psinone$ is to solve the modified Teukolsky equation at first order numerically, Eq.~\eqref{eq: first order modified teukolsky equation}.
Another alternative to compute the mode shift is to use the Green's function (as in the spectral decomposition above), but evaluated, with an appropriate regularisation, at the QNM frequency. This yields an integral representation for the mode shift, given by Eqs.~\eqref{eq: integral representation of chi1},\eqref{eq: integral representation of chi1 - regular part},\eqref{eq: integral representation of chi1 - regularized singular part} in App.~\ref{app: integral representation of mode shift}.

\section{Problem I: P{\"o}schl-Teller Potential}\label{sec: PT potential}

In this section, we apply the machinery laid down in the previous sections to a toy model featuring the Pöschl-Teller potential. Our aim is to compute the first and second order frequency shifts caused by a small change in width of the P{\"o}schl-Teller potential and compare them with their known values. We also explore the spectral decomposition of the first-order mode shift, showing that the QNM sum in \eqref{eq: implicit spectral decomposition} diverges in this context. %

Consider the $1+1$ wave equation
\begin{equation}
\label{RW-like equation}
    \left(-\partial_{tt} + \partial_{xx} - V(x,\zeta)\right) \psi(t,x,\zeta) = 0 \, ,
\end{equation}
where $x \in (-\infty,+\infty)$ plays the role of the tortoise coordinate. The majority of the formulae of this section apply to \eqref{RW-like equation} with a potential $V(x)$ with properties similar to the potential of the Regge-Wheeler (RW) equation \cite{Berti:2009kk}. Hence we will call \eqref{RW-like equation} a RW-like equation. 
For the computation of first and second order frequency shifts, as well as the first order mode shift, we specialize to the case of the P{\"o}schl-Teller potential
\begin{equation}
    V(x,\zeta) = \frac{1}{\cosh^2(\zeta x)} \, .
\end{equation}
Here $\zeta$ controls the width of the potential. Without loss of generality, we set the unperturbed width to $\zeta=1$. 

Quasinormal modes $\psi_n(\zeta)$ and frequencies $\omega_n(\zeta)$ of the P{\"o}schl-Teller potential are known analytically in closed form, as summarized in the next subsection. Hence, the frequency shifts $\omega_n^{(1)}$, $\omega_n^{(2)}$ can be computed exactly by differentiating $\omega_n(\zeta)$ with respect to $\zeta$ (and setting $\zeta=1$). 
In this section we show that we can recover these values 
using our perturbative formalism.

\subsection{QNMs of P{\"o}schl-Teller}\label{subsec: QNMs of Pöschl-Teller}

Following e.g.~\cite{Berti:2009kk}, we write a QNM solution of Eq.~\eqref{RW-like equation} as $\psi_n(t,x,\zeta)=e^{-i \omegan(\zeta)t}R_n(x,\zeta)$. The mode function $R_n$ %
is given by
\begin{equation}
\label{eq: PT QNM}
    R_n(y,\zeta) = (y(1-y))^{-i \omega_n(\zeta)/2\zeta}  {}_2 F_1(a_n,b_n,c_n;y) \, ,
\end{equation}
where $y = (1+e^{-2 \zeta x})^{-1}$,
\begin{eqnarray}
  a_n,b_n &=& \frac{1}{2} - \frac{i \omega_n(\zeta)}{\zeta} \pm \frac{1}{2}\sqrt{1-\frac{4}{\zeta^2}} , \\
  c_n &=& 1-\frac{i \omega_n(\zeta)}{\zeta} ,
\end{eqnarray}
and
\begin{equation}
\label{eq: PT frequency}
    \omega_{n}(\zeta) = \pm \sqrt{1-\zeta^2/4} - i \zeta ( n + 1/2) \, . 
\end{equation}
In what follows we define $\omega_{n}^{(0)} = \omega_{n}(\zeta=1)$, $R_n^{(0)}=R_n(\zeta=1)$ and $\psi_n^{(0)}=\psi_n(\zeta=1)$.

\subsection{Bilinear form for QNMs of Regge-Wheeler-like equations}\label{subsec: Pöschl-Teller bilinear form}

We now derive the bilinear form for RW-like equations. In this section and this section only, the operator $\odaggzero$ is defined as
\begin{equation}
\label{eq: RW-like operator}
    \odaggzero = -\partial_{tt} + \partial_{xx} - V(x,\zeta=0) \, ,
\end{equation}
and Latin indices are either $0$ or $1$.
To compute the conserved current $\pi^{a}$ we start from Eq.~\eqref{symplectic current} to find the Klein-Gordon-like current
\begin{equation}
    \pi^a[\psi_1,\psi_2] = (\psi_1 \partial_t \tilde{\psi}_2 - \tilde{\psi}_2 \partial_t \psi_1, \ \tilde{\psi}_2 \partial_x \psi_1 - \psi_1 \partial_x \tilde{\psi}_2) \, .
\end{equation}
As a byproduct, one also finds $\odaggzero = \mathcal{O}^{(0)}$.
From the current, noting that in this context the symmetry operator $\mathcal{J}$ simply reduces to $J_t:t \rightarrow-t$, follows the bilinear form
\begin{equation}
    \langle \! \langle \psi_1, \psi_2 \rangle \! \rangle = \int_{\mathcal{C}} \mathrm{d}x \left[ (J_t \psi_1) \partial_t \psi_2 - \psi_2 \partial_t (J_t \psi_1) \right] \, ,
\end{equation}
where $\mathcal{C}$ is some contour that regularizes the integral. We will turn to this issue later (Sec.~\ref{subsec: PT first order freq shift}).

The balance law for this problem is
\begin{equation}
    \frac{\mathrm{d}}{\mathrm{d}t}\langle \! \langle \psi_1, \psi_2 \rangle \! \rangle = \int_{\mathcal{C}} \mathrm{d}x \left ( \tilde{\psi}_2 \mathcal{O}^{(0)} \psi_1 - \psi_1 \odaggzero \tilde{\psi}_2 \right ) \, .
\end{equation}
The norm of a QNM of the RW-like %
equation is simply given by
\begin{equation}
    \langle \! \langle \psi_{n}^{(0)},\psi_{n}^{(0)} \rangle \! \rangle  = -2i\omega_{n}^{(0)} \int_{\mathcal{C}} \mathrm{d}x R_n^{(0)2}.
\end{equation}
Another useful identity is the following: given a scalar $\psi = R(x)e^{-i\omega t}$, one can show
\begin{equation}
\label{eq: general projection on a QNM - RW-like case}
    \langle \! \langle \psi_{n'}^{(0)}, \psi \rangle \! \rangle = -i (\omega + \omega^{(0)}_{n'})  \int_{\mathcal{C}} \mathrm{d}x (J_t\psi_{n'}^{(0)}) \psi \ .
\end{equation}

\subsection{Formulae for Schr{\"o}dinger-like perturbation theory}\label{subsec: Pöschl-Teller perturbation theory formulas}

Here, exploiting the identities of the last subsection, we list the main formulae for Schr{\"o}dinger-like perturbation theory.
We define 
\begin{equation}
    \odaggone= -V^{(1)}(x) = -\mathrm{d}V(x,\zeta)/\mathrm{d}\zeta |_{\zeta=\zeta_0} \, , 
\end{equation}
where $\zeta_0$ is some fixed value. For P{\"o}schl-Teller we choose $\zeta_0=1$, with the idea that the perturbation to the potential represents a small change in the width of the P{\"o}schl-Teller potential. 

For the first order frequency shift, we find
\begin{equation}
    \label{eq: first order freq shift PT}
    \omega_{n}^{(1)} = i \frac{\int_{\mathcal{C}}\mathrm{d}x J_t \psi^{(0)}_n \odaggone \psi_n^{(0)}}{\langle \! \langle \psi_{n}^{(0)}, \psi_n^{(0)} \rangle \! \rangle} = \frac{\int_{\mathcal{C}}\mathrm{d}x R^{(0)2}_n V^{(1)}}{2 \omega_{n}^{(0)} \int_{\mathcal{C}}\mathrm{d}x R^{(0)2}_n} \, .
\end{equation}
Similarly, the projection coefficients ($n' \neq n$ here) read
\begin{eqnarray}
\label{eq: projection coefficients PT}
  \frac{\langle \! \langle \psi_{n'}^{(0)}, \psi_n^{(1)} \rangle \! \rangle}{\langle \! \langle \psi_{n'}^{(0)}, \psi_{n'}^{(0)} \rangle \! \rangle} &=& \frac{i}{\Delta \omega^{(0)}_{nn'}} \frac{\int_{\mathcal{C}} \mathrm{d}x J_t \psi^{(0)}_{n'} \mathcal{O}^{(1)} \psi_n^{(0)}}{\langle \! \langle \psi_{n'}^{(0)}, \psi_{n'}^{(0)} \rangle \! \rangle} \nonumber \\
  &=& \frac{e^{-i \Delta\omega_{nn'}^{(0)}t }}{\Delta \omega^{(0)}_{nn'}} \frac{\int_{\mathcal{C}}\mathrm{d}x R^{(0)}_{n'} V^{(1)} R_n^{(0)}  }{2 \omega^{(0)}_{n'} \int_{\mathcal{C}}\mathrm{d}x R^{(0)2}_{n'}} .
\end{eqnarray}
Whereas, the equation for the second order frequency shift is 
\begin{eqnarray}
\label{eq: PT second order freq shift with R1}
  \omega_n^{(2)} &=& -i\frac{\int\mathrm{d}x R_n^{(0)2}V^{(2)}}{\langle \! \langle \psi_{n}^{(0)}, \psi_n^{(0)} \rangle \! \rangle} - 2i\frac{\int\mathrm{d}x R_n^{(0)}V^{(1)}R_n^{(1)}}{\langle \! \langle \psi_{n}^{(0)}, \psi_n^{(0)} \rangle \! \rangle} \nonumber \\
  && + 4 i \omega_n^{(0)}\omega_n^{(1)}\frac{\int\mathrm{d}x R_n^{(0)}R_n^{(1)}}{\langle \! \langle \psi_{n}^{(0)}, \psi_n^{(0)} \rangle \! \rangle} - \frac{\omega_n^{(1)2}}{\omega_n^{(0)}}.
\end{eqnarray}

\subsection{Computing first order frequency shifts for the P{\"o}schl-Teller case}\label{subsec: PT first order freq shift}

As discussed in Sec.~\ref{subsec: QNMs of Pöschl-Teller}, the quasinormal frequency $\omega_n(\zeta)$ of the P{\"o}schl-Teller potential is known analytically for every $\zeta$. Hence, by differentiating \eqref{eq: PT frequency} with respect to $\zeta$ and setting $\zeta=1$ we find the frequency shift
\begin{equation}\label{eq:PT_omega1_true}
    \omega_n^{(1)} = -\frac{1}{2 \sqrt{3}} - i (n + 1/2).
\end{equation}
We will now show that Eq.~\eqref{eq: first order freq shift PT} for $\omega_n^{(1)}$ yields the same result. To this end, we need to compute the integrals appearing in Eq.~\eqref{eq: first order freq shift PT}. Reference \cite{Leung_1998} encountered the same integrals when computing $\omega_n^{(1)}$ exploiting a different product. To evaluate them, they (i) replace $\omega_n^{(0)}$ with a generic $\omega$ in the upper half plane, 
with the idea of performing the following integrals on the real line 
\begin{eqnarray}
  && I(\omega) = \int_{-\infty}^{+\infty} \mathrm{d}x R^{(0)2}_n(x,\omega) V^{(1)}(x), \\
  && J(\omega) = \int_{-\infty}^{+\infty} \mathrm{d}x R^{(0)2}_n(x,\omega) \, ;
\end{eqnarray}
(ii) as the integrals generally diverge for $\omega=\omega_n^{(0)}$, they find a closed form solution for them in the case $n=0,1$ in terms of special functions and then by analytic continuation of $I(\omega),J(\omega)$ compute $I(\omega_n^{(0)}),J(\omega_n^{(0)})$. We here do the same, showing that their procedure can be carried out for generic $n$ by expressing P{\"o}schl-Teller QNMs via hypergeometric functions \eqref{eq: PT QNM}, and by exploiting the series representation
\begin{equation}
    {}_2F_1(a,b,c,;z)=\sum^{\infty}_{k=0} \frac{(a)_k (b)_k}{(c)_k} \frac{z^k}{k!} \, ,
\end{equation}
where $(q)_k$ is the Pochhammer symbol.
Using this power series and the Cauchy product, we can write 
\begin{equation}
    ({}_2F_1(a,b,c,;z))^2 = \sum^{\infty}_{j=0} K_j z^j \, ,
\end{equation}
where
\begin{equation}
    K_j = \sum_{i=0}^{j} \frac{(a)_i (b)_i}{(c)_i} \frac{(a)_{j-i} (b)_{j-i}}{(c)_{j-i}} \, .
\end{equation}
Hence, using $R^{(0)}_n(x) = R_n(y(x),\zeta=1)$ and
\begin{equation}
    V^{(1)}(x) = -2x \text{sech}^2x \tanh x ,
\end{equation}
we write $I,J$ as
\begin{eqnarray}
  I(\omega) =- \sum_{k=0}^{\infty} 2^{1+2i \omega} K_j(\omega) \int_{-\infty}^{+\infty} \mathrm{d}x && \big[ x \tanh{x} (\text{sech}^2x)^{1-i \omega} \nonumber \\
  && \times (1+e^{-2x})^{-j} \big],
\end{eqnarray}
\begin{equation}
 J(\omega) = \sum_{k=0}^{\infty} 4^{i \omega} K_j(\omega) \int_{-\infty}^{+\infty} \mathrm{d}x (\text{sech}^2x)^{-i \omega} (1+e^{-2x})^{-j} .
\end{equation}
These integrals can be solved in terms of incomplete beta functions and generalized hypergeometric functions. One can then compute $I, J$ for any $\omega_n^{(0)}$ via analytic continuation of the featured special functions. 

In terms of $I, J$, the frequency shift \eqref{eq: first order freq shift PT} reads
\begin{equation}
\label{operative freq shift PT}
    \omega_n^{(1)} = \frac{I(\omega_n^{(0)})}{2 \omega_n^{(0)} J(\omega_n^{(0)})}\, .
\end{equation}
Table \ref{tab: first order freq shift PT - agreement} shows the exact agreement between the true frequency shifts \eqref{eq:PT_omega1_true} and Eq.~\eqref{operative freq shift PT}.
\begin{table}
  \caption{\label{tab: first order freq shift PT - agreement}Exact (second column) frequency shift to the $n$-th QNM of the P{\"o}schl-Teller potential compared to its computation using formula \eqref{operative freq shift PT}, based on the bilinear form (third column). Here we show the first few digits, but we have used Mathematica to check that the two quantities are numerically equal%
    .}
    \begin{ruledtabular}
    \renewcommand{\arraystretch}{1.2}
    \begin{tabular}{ccc}
        $n$ & $ \omega_n^{(1)} = -(2 \sqrt{3})^{-1} - i (n + 1/2)  $ & $ \omega_n^{(1)} = \dfrac{I(\omega_n^{(0)})}{2 \omega_n^{(0)} J(\omega_n^{(0)})} $ \\[6pt]
        \colrule
        0 & $-0.288675 -i \ 0.5  $ & $ -0.288675 -i \ 0.500000 $\\
        1 & $-0.288675 - i \ 1.5 $ & $ -0.288675 - i \ 1.500000 $ \\
        2 & $ -0.288675 - i \ 2.5$ & $ -0.288675 - i \ 2.500000$ \\
        5 & $ -0.288675 - i \ 5.5$ & $-0.288675 - i \ 5.500000 $ \\
        10 & $ -0.288675 - i \ 10.5$ & $ -0.288675 - i \ 10.500000$ \\
        15 & $ -0.288675 - i \ 15.5$ & $-0.288675 - i \ 15.500000 $ \\
        30 & $ -0.288675 - i \ 30.5$ & $ -0.288675 - i \ 30.500000$ \\
        40 & $ -0.288675 - i \ 40.5$ & $ -0.288675 - i \ 40.500000$ \\
        50 & $ -0.288675 - i \ 50.5$ & $ -0.288675 - i \ 50.500000$ \\
        60 & $ -0.288675 - i \ 60.5$ & $ -0.288675 - i \ 60.500000$ \\
        70 & $ -0.288675 - i \ 70.5$ & $ -0.288675 - i \ 70.500000$ \\
    \end{tabular}
    \end{ruledtabular}
\end{table}

\subsection{Divergence of QNM sum in mode shift spectral decomposition}\label{subsec: PT mode shift}

In this section, as anticipated in Sec.~\ref{sec:Modeshifts}, we show that the QNM sum in the spectral decomposition of the mode shift \eqref{eq: implicit spectral decomposition} is divergent. 

The projection coefficients in the RW-like case are given by \eqref{eq: projection coefficients PT}. We then need to compute the sum of integrals involving the secular term in \eqref{eq: implicit spectral decomposition}. In this context we find that the projection of the secular term is zero,
\begin{eqnarray}
\label{eq: RW-like simplification of the secular term}
  &&\surf t^a \ps \psinpzero \odaggzero (t \psinzero) \nonumber\\
  &&\quad = 2 i \omega_{n}^{(0)} \int \mathrm{d}x (J_t \psi_{n'}^{(0)}) \psi_{n}^{(0)} \nonumber \\
  &&\quad = -\frac{2\omega_n^{(0)}}{\omega_{n}^{(0)}+ \omega_{n'}^{(0)}} \langle \! \langle \psi_{n'}^{(0)}, \psi_{n}^{(0)}  \rangle \! \rangle
  = 0,
\end{eqnarray}
where we have used Eqs.~\eqref{eq: RW-like operator}, \eqref{eq: general projection on a QNM - RW-like case}, and in the last step orthogonality. 
From the definition of the mode shift 
\begin{equation}
    \psi_{n}^{(1)}(t,x) = -i \omega^{(1)}_n t \psi_n^{(0)}(t,x) + e^{-i \omega_n^{(0)}t}R^{(1)}_n(x),
\end{equation}
using \eqref{eq: spatial mode shift in terms of QNM sum} and \eqref{eq: RW-like simplification of the secular term} we get 
\begin{equation}
\label{eq: PT QNM sum for mode shift}
    R_{\text{QNM}}^{(1)}(x) = \sum_{n'\neq n}\frac{-i}{\Delta\omega_{nn'}^{(0)}}\frac{\int_{-\infty}^{\infty} R_{n'}^{(0)}V^{(1)}R_n^{(0)}\mathrm{d}x'}{\langle\!\langle \psi_{n'}^{(0)},\psi_{n'}^{(0)}\rangle \!\rangle}R_{n'}^{(0)}(x).
\end{equation}

\begin{figure}[th]
    \centering
    \includegraphics[width=\linewidth]{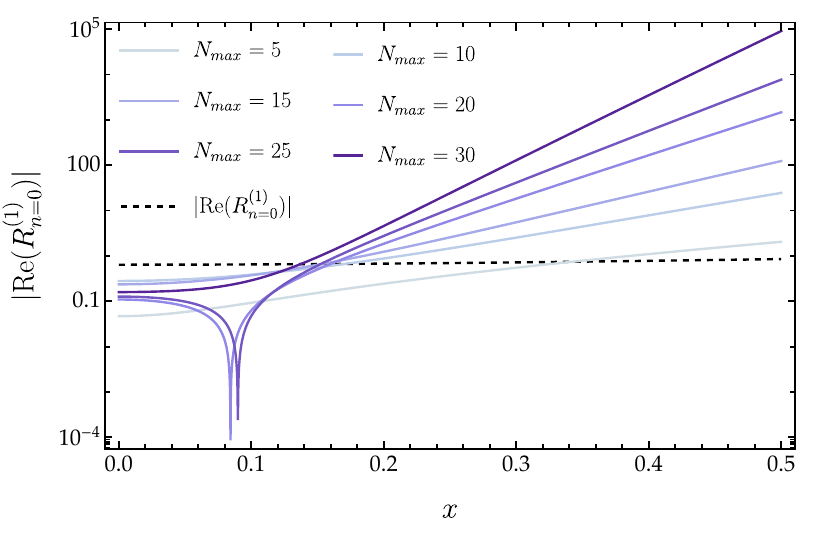}
    \caption{%
    Real part of the QNM contribution to the mode shift $R^{(1)}_{n=0}(x)$, Eq.~\eqref{eq: PT QNM sum for mode shift}. The sum over QNMs does not converge to the true mode shift of the fundamental mode of the PT spectrum (black dashed line) in the interval $x \in [0,0.5]$. Solid lines represent partial QNM sums including an increasing number $N_{max}$ of overtones and $N_{max}$ mirror modes. The true mode shift only appears constant due to the plot scale. }
    \label{fig: QNM sum divergence}
\end{figure}

We conclude by showing that the QNM contribution~\eqref{eq: PT QNM sum for mode shift} is divergent in the P{\"o}schl-Teller problem. We compute the projection coefficients in Eq.~\eqref{eq: PT QNM sum for mode shift} using the same techniques discussed in the computation of the first order frequency shift, and compute partial QNM sums. As seen in Fig.~\ref{fig: QNM sum divergence}, the sum blows up the more overtones we include. 
We expect this divergence to be regularized by the continuous-spectrum contribution, Eq.~\eqref{eq: continuous spectral part of modeshift}. Furthermore, this strongly suggests that the divergence of the QNM sum in the spectral decomposition of the mode shift \eqref{eq: implicit spectral decomposition} is a general feature of QNMs.

\subsection{Second order frequency shift of the fundamental mode}\label{subsec: PT mode shift and second order frequency shift}

To conclude the study of the P{\"o}schl-Teller potential, we test the formula for the second order frequency shift by using the knowledge of the full solution $R_n(x,\zeta)$ in this setting.

We can simply compute $R_n^{(1)}(x)$ by differentiating $R_n(x,\zeta)$ with respect to $\zeta$. For $n=0$ we have
\begin{eqnarray}
  R^{(0)}_0(x) &=& 2^{i \omega^{(0)}_0}(\cosh x)^{i \omega^{(0)}_0}, \\
  R^{(1)}_0(x) &=& 2^{i \omega^{(0)}_0} i \omega^{(0)}_0 (\cosh x)^{i \omega^{(0)}_0 - 1}\big(  x \sinh x \nonumber \\
  && - (1-\omega_0^{(1)}/\omega^{(0)}_0) \cosh x \log (2\cosh x)\big),
\end{eqnarray}
and also
\begin{equation}
    V^{(2)}(x) = 2x^2(\cosh(2x) - 2)\mathrm{sech}^4 x.
\end{equation}
Plugging these terms into the integrals of Eq.~\eqref{eq: PT second order freq shift with R1} would lead to the usual divergences due to the QNMs blowing up on the real axis. 
To circumvent this issue, as we have discussed in Sec.~\ref{subsec: PT first order freq shift}, we replace $\omega_n^{(0)}$ by a generic $\omega$ in the upper half plane. Then it is possible to show that all the integrals admit a closed form solution in terms of Beta, Gamma and Digamma functions. By analytically continuing the special functions to $\omega=\omega_0^{(0)}$, we find
\begin{equation}
  \omega^{(2)}_0 = -0.38490017945975 + O(10^{-15})i ,
\end{equation}
exactly matching the value of the known $\omega^{(2)}_0$,
\begin{equation}
  \left. \frac{\mathrm{d}^2\omega(\zeta)}{\mathrm{d}\zeta^2} \right|_{\zeta=1} = -\frac{2}{3 \sqrt{3}} \simeq -0.38490017945975 .
\end{equation}
This confirms the validity of our formula for the second order frequency shift, once the mode shift is obtained.

\section{Problem II: slowly spinning black hole}\label{sec: SchwToKerr}

We now show how to use the formalism we developed to compute the frequency shift at first and second order when dealing with Teukolsky-like equations. To do so, we consider a slowly rotating Kerr BH and treat it as a deformed Schwarzschild background. Due to the structure of the Teukolsky operator, the formulae to compute the frequency shifts are much more complicated compared to those of RW-like operators. This is why the study of a slowly spinning BH is needed to further assess the effectiveness of our formalism.

\begin{widetext}
The Teukolsky operator for Kerr reads \cite{Teukolsky:1973ha}
\begin{eqnarray}
\label{eq: TeukolskyKerr}
 _s\mathcal{O} &=& \frac{1}{\Sigma} \left\{ \left[\frac{(r^2+a^2)^2}{\Delta} - a^2 \sin^2 \theta \right] \partial_{tt} + \frac{4Mar}{\Delta}\partial_{t\phi} \nonumber \right. + \left[ \frac{a^2}{\Delta} - \frac{1}{\sin^2\theta} \right] \partial_{\phi \phi} - \Delta^{-s} \partial_r (\Delta^{s+1}\partial_r) \nonumber \\
  && - \frac{1}{\sin \theta} \partial_{\theta}(\sin\theta \partial_{\theta}) -2s \left[ a\frac{r-M}{\Delta}+ \frac{i \cos\theta}{\sin^2\theta} \right] \partial_{\phi} \left. -2s \left[ M \frac{r^2 - a^2}{\Delta} - r - ia\cos\theta \right]\partial_t + s^2 \cot^2 \theta - s \right\}.
\end{eqnarray}
We define $_s\ozero$ setting $a=0$, %
\begin{eqnarray}
\label{eq: TeukolskySchwarzschild}
  _s\ozero &=& \frac{1}{f} \partial_{tt} - \frac{1}{r^2 \sin^2\theta}\partial_{\phi \phi} -\frac{1}{r^2} \left( r^2 f \right) ^{-s} \partial_r \left[\left( r^2f \right)^{s+1}\partial_r \right] \nonumber - \frac{1}{r^2 \sin \theta} \partial_{\theta}(\sin\theta \partial_{\theta}) - \frac{2s}{r^2} \frac{i \cos\theta}{\sin^2\theta} \partial_{\phi} \nonumber \\
  && -\frac{2s}{r^2} \left( \frac{M}{f} - r \right) \partial_t + \frac{1}{r^2}(s^2 \cot^2 \theta - s) ,
\end{eqnarray}
where $f(r) = 1 - 2M/r$. 
Also, writing $a = \zeta M$, $\zeta \ll 1$, expanding \eqref{eq: TeukolskyKerr} and neglecting terms of order $\zeta^2$ and higher we find, fixing $s=\pm2$,  $\mathcal{O}_{(1)}$ and $\odaggone$
\begin{equation}
\label{eq: perturbedOschw}
    \oone =  \frac{4M^2}{r^3 f} \partial_{t \phi} - \frac{4M}{r^3 f} (1-\frac{M}{r}) \partial_{\phi} + \frac{4M}{r^2}i\cos\theta \partial_t ,
\end{equation}
\begin{equation}
\label{perturbedOdaggSchw}
    \odaggone = \frac{4M^2}{r^3 f} \partial_{t \phi} + \frac{4M}{r^3 f} (1-\frac{M}{r}) \partial_{\phi} - \frac{4M}{r^2}i\cos\theta \partial_t .
\end{equation}
Similarly, for $\odagg_{(2)}$ we have
\begin{eqnarray}
  \odagg_{(2)} &=& - \frac{2 M^2 \cos^2\theta}{r^2} \odaggzero + \frac{M^2}{r^2} \left[ - \frac{16 M}{r^2 f^2} \left( 1 - \frac{M}{r} \right) \partial_t + \left( \frac{2}{f^2} \left( 1 - \frac{4 M}{r} \right) - 2 \sin^2 \theta \right) \partial_{tt} + \frac{2}{r^2 f} \partial_{\phi \phi} - 2 \partial_{rr}   \right].
\end{eqnarray}
\end{widetext}
A QNM of the Kerr spacetime satisfies $\odagg(a) \psin(a) = 0$, which, expanding up to second order in $\zeta=a/M$ yields the first and second order equations
\begin{eqnarray}
  && \odaggzero \psinone = - \odaggone \psinzero, \\
  && \odaggzero \psin^{(2)} = - \odagg_{(2)} \psinzero - 2 \odaggone \psinone.
\end{eqnarray}
We start by writing the formula for the first-order frequency shift $\omegan^{(1)}$ \eqref{first order frequency shift} in this setting, explicitly in BL coordinates (in the following $\Delta, \Sigma$ are always evaluated at $a=0$). We have
\begin{eqnarray}
  \odaggone \psinzero &=& e^{-i \omegan^{(0)}t} e^{i m \phi} \frac{1}{r^2} S_{lm}^{(0)} R_{ln}^{(0)} \Big[ -4 M \omegan^{(0)} \cos\theta \nonumber \\
  && + \frac{4 M m}{r f} \lp M \omegan^{(0)} + i \lp 1-\frac{M}{r} \rp \rp \Big] .
\end{eqnarray}
Then, using Eq.~\eqref{reflection in BL coordinates}
alongside $\mathrm{d}\Sigma_a t^a = \mathrm{d}r\mathrm{d}\theta\mathrm{d}\phi \Sigma \sin\theta$ and the normalisations $ 2 \pi \int_0^{ \pi} \mathrm{d}\theta \sin\theta S_{lm}^{(0)} S_{l'm}^{(0)} = \delta_{ll'}$ and $ \int_0^{2 \pi} \mathrm{d}\phi e^{i(m-m')\phi}=2 \pi \delta_{mm'}$ , 
we find
\begin{equation}
    \label{BL coord Freq shift 1}
    \omegan^{(1)} =4 i M^{4/3}\frac{\int \d r \Delta^{-2} R^{(0)2}_{\boldsymbol n} \lp g_{\boldsymbol n}(r)+\zeta^{(1)}_{\boldsymbol n} \rp }{\normQNM},
\end{equation}
where the explicit expression for the norm of the selected Schwarzschild QNM is given by setting $a=0$ in Eq.~(55) of \cite{Green:2022htq}, and reads
\begin{equation}
    \label{Norm QNM Schwarzschild}
    \normQNM = 4 i M^{4/3} \int \d r \Delta^{-2} h_{\boldsymbol n}(r) R^{(0)2}_{\boldsymbol n}.
\end{equation}
The radial functions $h_{\boldsymbol n}, g_{\boldsymbol n}$ are
\begin{eqnarray}
  && g_{\boldsymbol n}(r) = \frac{4 M m}{r f} \lp M \omegan^{(0)} + i \lp 1-\frac{M}{r} \rp \rp, \\
  && h_{\boldsymbol n}(r)= -\frac{r}{f} \lp -2 \omegan^{(0)} r + 4 i \lp 1-\frac{3 M}{r} \rp \rp,
\end{eqnarray}
and $\zeta^{(1)}_{\boldsymbol n}$ is the first order shift to the separation variable, given by
\begin{equation}
\label{lambda shift}
    \zeta^{(1)}_{\boldsymbol n} = - 8 \pi M \omegan^{(0)} \int^{\pi}_0 \d \theta \sin\theta \cos\theta S^{(0)2}_{\boldsymbol n}.
\end{equation}

We computed the radial integral for $\omegan^{(1)}$ along the contour described in Sec.~\ref{sec: Product} for various values of $l$ and $n$, using Leaver's series expansion for $R^{(0)}_{\boldsymbol n}$ \cite{Leaver:1985ax}, modified to place the branch-cut as shown in Fig.~\ref{plot: Leaver's contour}. Leaver's expansion with the rotated branch-cut reads ($M$=1 here)
\begin{eqnarray}
\label{Leaver's series}
  R_{\boldsymbol n}^{(0)} &=& e^{i \omegan^{(0)}r } ( i r )^{1+4 i \omegan^{(0)}} ( i (r-2) )^{2-2i \omegan^{(0)}} \nonumber \\
  && \times (2 i)^{ -3 - 2 i \omegan^{(0)}} \sum_{k=0}^{\infty} d_k f^k(r).
\end{eqnarray}
Note that this expansion is convergent along the chosen contour.
The computation allows us to numerically verify the equality
\begin{equation}
    \omegan^{(1)} = \left. \frac{\d \omegan(a)}{\d a} \right|_{a=0}.
\end{equation}
Indeed, Table~\ref{tab: first order freq shift} shows that $\frac{\omegan(a)-\omegan(0)}{a}$ converges linearly to $\omegan^{(1)}$.

\begin{table}
    \caption{\label{tab: first order freq shift}Here $\delta \omega(a) = (\omega(a)-\omega(0))/a$, where $\omega(a)$ is the Kerr spectrum. The claim is that: $\omegan^{(1)} = \d \omega(a) / \d a |_{a=0}$. The table shows that, as $a \rightarrow 0$, the relative error between $\delta \omega(a)$ and $\omegan^{(1)}$ goes to zero.}
    \begin{ruledtabular}
    \renewcommand{\arraystretch}{1.2}
    \begin{tabular}{ccc}
        $a$ & $1 - \mathrm{Re}(\delta\omega)/\mathrm{Re}(\omegan^{(1)})$ & $1 - \mathrm{Im}(\delta\omega)/\mathrm{Im}(\omegan^{(1)})$ \\
        \colrule
        $10^{-3}$ & $5.71 \cdot 10^{-4}$ & $2.63 \cdot 10^{-3}$ \\
        $10^{-4}$ & $6.23 \cdot 10^{-5}$ & $2.66 \cdot 10^{-4}$ \\
        $10^{-5}$ & $1.99 \cdot 10^{-5}$ & $7.94 \cdot 10^{-5}$ \\
    \end{tabular}
    \end{ruledtabular}
\end{table}

We now turn to discuss the computation of the modal shift $\psinone$ and second order frequency shift. In this simple setting, due to axisymmetry of the perturbed spacetime, the ansatz for $\psinone$ reduces to
\begin{equation}
    \psinone = -i \omegan^{(1)}t \psinzero + e^{-i \omegan^{(0)}t} e^{i m \phi} \lp S_{\boldsymbol n}^{(0)}R_{\boldsymbol n}^{(1)} + S_{\boldsymbol n}^{(1)}R_{\boldsymbol n}^{(0)} \rp,
\end{equation}
where $S_{\boldsymbol n}^{(1)}$ can be easily obtained expanding in $a$ the spheroidal harmonics $S_{\boldsymbol n}(a)$. 
To verify the validity of our second-order frequency shift formula, we approximate $R_{\boldsymbol n}^{(1)}(a)$ as
\begin{equation}
R_{\boldsymbol n}^{(1)}(a) = \frac{R_{\boldsymbol n}(a)-R_{\boldsymbol n}^{(0)}}{a},
\end{equation}
with $R_{\boldsymbol n}(a)$ given by Leaver's series.
$R^{(1)}_{\boldsymbol n}(a)$ can then be used to evaluate $\omegan^{(2)}$ through Eq.~\eqref{second order shift} or \eqref{eq: appendix - effective second order freq shift formula}. In BL coordinates, Eq.~\eqref{eq: appendix - effective second order freq shift formula} reads
\begin{eqnarray}
  \omegan^{(2)} &=& -\frac{4 i M^{4/3}}{\normQNM}\bigg[ 2 M^2\int_{\mathcal{C}} \d r \Delta^{-2} R_{\boldsymbol n}^{(0)} \frac{\d^2 R_{\boldsymbol n}^{(0)}}{\d r^2} \nonumber \\
  && \quad + 2 \int_{\mathcal{C}} \d r \Delta^{-2} R_{\boldsymbol n}^{(0)} R_{\boldsymbol n}^{(1)} (\omegan^{(1)} h_{\boldsymbol n} - g_{\boldsymbol n} -\zeta_{\boldsymbol n}^{(1)}) \nonumber \\
  && \quad + \int_{\mathcal{C}} \d r \Delta^{-2} R_{\boldsymbol n}^{(0)2}\bigg( J_{\boldsymbol n} + K_{\boldsymbol n} + L_{\boldsymbol n} +  p_{\boldsymbol n} \nonumber \\
  && \quad + \omegan^{(1)2} \frac{2 r^2}{f} - 2 M \omegan^{(1)}\frac{4 m M}{r f} \bigg) \bigg],
\end{eqnarray}
where
\begin{eqnarray}
  p_{\boldsymbol n}(r) &=& \frac{2 m^2 M^2}{r^2 f} - \frac{16 i M^3 \omegan^{(0)}}{r^2 f^2} \nonumber \\ && \times
  \left( 1-\frac{M}{r}-\frac{i}{2}M \omegan^{(0)} \right), \\
  J_{\boldsymbol n} &=& 16 \pi M \omegan^{(0)}\int \mathrm{d}\theta \sin\theta \cos\theta S_{\boldsymbol n}^{(0)}S_{\boldsymbol n}^{(1)}, \\
  K_{\boldsymbol n} &=& 4 \pi M^2 \omegan^{(0)2}\int \mathrm{d}\theta \sin\theta \cos^2\theta S_{\boldsymbol n}^{(0)2}, \\
  L_{\boldsymbol n} &=& 16 \pi M\omegan^{(1)} \int \mathrm{d}\theta \sin\theta \cos\theta S_{\boldsymbol n}^{(0)2}.
\end{eqnarray}
The resulting second order frequency shift depends on the perturbative parameter $a$, $\omegan^{(2)} \equiv  \omegan^{(2)}(a)$, as we used $R_{\boldsymbol n}^{(1)}(a)$. The comparison of $\omegan^{(2)}(a)$ with $\d^2 \omegan(a)/\d a^2 |_{a=0}$ is shown in Table \ref{tab: second order freq shift}.
As expected,  the relative error between the two quantities goes to zero as $a \rightarrow 0$. 

\begin{table}
    \caption{\label{tab: second order freq shift}Here, $\delta^2 \omega$ is the second derivative of $\omega(a)$ computed using finite differences with $a=10^{-4}$ and an $8$-point stencil. The claim is that: $\omegan^{(2)}(a) = \d^2 \omega(a) / \d a^2 |_{a=0}$ as $a \rightarrow 0$. The table shows that the relative error between $\delta^2 \omega$ and $\omegan^{(2)}(a)$ goes to zero with the spin $a$, as expected. %
    }
    \begin{ruledtabular}
    \renewcommand{\arraystretch}{1.2}
    \begin{tabular}{ccc}
        $a$ & $1 - \mathrm{Re}(\delta^2\omega)/\mathrm{Re}(\omegan^{(2)}(a))$ & $1 - \mathrm{Im}(\delta^2\omega)/\mathrm{Im}(\omegan^{(2)}(a))$ \\
        \colrule
        $10^{-2}$ & $1.21 \cdot 10^{-3}$ & $2.52 \cdot 10^{-2}$ \\
        $10^{-3}$ & $1.25 \cdot 10^{-4}$ & $2.39 \cdot 10^{-3}$ \\
        $10^{-4}$ & $1.26 \cdot 10^{-5}$ & $2.37 \cdot 10^{-4}$ \\
    \end{tabular}
    \end{ruledtabular}
\end{table}
As in the case of the P\"oschl-Teller problem, we also computed the first few terms of the QNM sum contribution to the mode shift $\psinone$, appearing in Eq.~\eqref{eq: implicit spectral decomposition}. The coefficients are harder to compute beyond the first few overtones, but also in this case we find indication that the sum does not converge.

\section{Conclusions}
\label{sec:conclusions}

In this work, we combined the modified Teukolsky equation with the orthogonal QNM bilinear form to build a systematic expansion analogous to time-independent perturbation theory in quantum mechanics. Our main motivation is the problem of QNM spectral shifts induced by bGR theories or environmental effects. Our framework gives rise to formulas for frequency shifts at any order, in terms of lower-order mode shifts. We are also able to write down projection coefficients appearing in mode shifts. We further clarified the relationship between our bilinear form and other products based on the Sturm-Liouville form of the decoupled Teukolsky equations; namely, our projection coefficients are expressed as matrix elements using this product.

The main obstacle that arises in comparison to quantum mechanics is due to the lack of a spectral theorem, and hence the incompleteness of QNMs. This appears for the first time in the calculation of the leading order mode shift. Using Green's function techniques, we obtained its formal decomposition in terms of projections onto unperturbed QNMs, as well as continuum-spectrum components in the form of integrals over the branch cut and high-frequency arc in the complex-$\omega$ plane. The mode projections themselves take the natural form $\langle\!\langle \psi_{n'}^{(0)}, \psi_n^{(1)} \rangle\!\rangle $ expected using an orthogonal projector, however with an additional sum over projections of the secular term, which survives because the eigenvalue problem is nonlinear [see Eq.~\eqref{eq: implicit spectral decomposition}]. We expressed the projection terms as matrix elements of the modified-Teukolsky-equation source with respect to the Sturm-Liouville product. In both worked examples, we found numerical evidence that the QNM sum by itself diverges---a symptom of the lack of self-adjointness.

We identified two alternative methods to compute the mode shift that do not rely on possibly-ill-defined expansions. One can either directly compute the mode shift by numerically solving the modified Teukolsky equation (on the complex path that assures convergence of any integrals one wants to perform), or through the integral representation derived using Green's function techniques in App.~\ref{app: integral representation of mode shift}. %
As the calculation of higher order frequency shifts requires the knowledge of lower order mode shifts, we showed that, once $\psinone$ is at hand, the formula for $\omegan^{(2)}$ \eqref{second order shift} reproduces the known shift for both examples.

Going forwards, it would be interesting to extend our higher-order formulas to the degenerate case \cite{Li:2023ulk}, and further investigate the role of QNM incompleteness, particularly in the spectral decomposition of the first-order mode shift. To make our formal decomposition more practical, one would need to reorganize the decomposition so that the individual contributions (QNMs, continuum) are separately finite. %
Similar taming of divergent integrals was achieved in recent work on ringdown, using causality conditions to decompose the BH Green's function~\cite{DeAmicis:2025xuh,Kuntz:2025gdq,Arnaudo:2025uos,Su:2026fvj,Arnaudo:2026tcy,DeAmicis:2026wqd}. If they could be adapted, similar techniques might enable the continuous spectrum to be more fully incorporated into our perturbative framework.
One might then be able to assess if the continuum piece $\psi_{\boldsymbol{n},\gamma}^{(1)}$ and any unperturbed QNM are orthogonal to each other.

Our work also opens the possibility to explore various physical applications. %
Boson clouds offer an attractive testing ground in this respect. We expect the shift to a boson cloud's quasibound state to be likewise given by a mode sum and a continuum piece, $\psi_{\boldsymbol{n}}^{(1)} = \psi_{\boldsymbol{n},\text{modes}}^{(1)} + \psi_{\boldsymbol{n},\gamma}^{(1)}$. The connection with the hydrogen atom bound states in the non-relativistic limit suggests that the sum over quasibound states should converge in this case.
We expect our perturbative framework to be especially relevant in the boson cloud setting, as gravitational atoms with marginally unstable states could be sensitive to higher order spectral corrections~\cite{Guo:2026ToAppear}.

\begin{acknowledgments}
We thank P.~Arnaudo, E.~Cannizzaro, J.~Carballo, G.~D'Addario, S.~Hollands, T.~Sotiriou, and P.~Zimmerman  %
for helpful discussions.
LS acknowledges support from the UKRI Horizon guarantee funding (project no. EP/Y023706/1). LS is also supported by a University of Nottingham Anne McLaren Fellowship. 
S.R.G.~is supported by a UKRI Future Leaders Fellowship (grant number MR/Y018060/1).
\end{acknowledgments}

\appendix

\section{Details of second order frequency shift formula}
\label{app: second order details}

The goal of this section is to first show that the formula for the second order frequency shift \eqref{second order shift} is actually time independent, as it should. Indeed, by simply staring at \eqref{second order shift}, it may look that there must be a dependence on time, as $\psinone \notin \text{Ker}\odaggzero$ and hence $\projnone$ is not time independent. However, another dependence on time is carried by the integral featuring $\mathcal{S}_{(1)}\psinone$. We now show that the time dependence cancels out and we are left with a number. 

Consider the term of \eqref{second order shift} featuring $\projnone$. Using \eqref{eq: mode shift onto same mode}, we have:
\begin{equation}
\label{eq: time dependence in projnone}
    -2 \omegan^{(1)} \frac{\projnone}{\normQNM} = 2 i \omegan^{(1)2}t - 2 \omegan^{(1)}\frac{c}{\normQNM}.
\end{equation}
Consider now the integral featuring $\mathcal{S}_{(1)}\psinone$ in \eqref{second order shift} and plug in the ansatz
\begin{equation}
\label{eq: ansatz psione}
    \psinone=-i \omegan^{(1)}t \psinzero + e^{-i \omegan^{(0)}t}\chi^{(1)}(r,\theta,\phi).
\end{equation}
It is then clear that the time dependence would come from $\mathcal{S}_{(1)}(-i \omegan^{(1)}t \psinzero)$.
By exploiting the first order frequency shift formula \eqref{first order frequency shift} one has:
\begin{eqnarray}
  && 2i \frac{\surf t^a \ps \psinzero \mathcal{S}_{(1)}(-i \omegan^{(1)}t \psinzero)}{\normQNM}  =-2 i \omegan^{(1)2}t + \nonumber \\
  && 2\omegan^{(1)} \frac{\surf t^a \ps \psinzero \left[\mathcal{S}_{(1)}(t \psinzero)-t\mathcal{S}_{(1)}\psinzero\right]}{\normQNM}.
\end{eqnarray}
We have found the presence of $-2 i \omegan^{(1)2}t$, that cancels with the same term appearing in \eqref{eq: time dependence in projnone}. To conclude the proof we need to show that the integral featuring the term in square brackets is time independent. By construction, $\mathcal{S}_{(1)}$ is a linear differential operator built out of a finite number of time and spatial derivatives with purely spatial coefficients. For such an operator, it's easy to see that the integral must indeed be time independent. For example, in a realistic scenario where $\mathcal{S}_{(1)}$ has only time derivatives up to second order, and can then be written as
\begin{equation}
\label{eq: decomposition of S1}
    \mathcal{S}_{(1)} = \beta \partial_{tt}+\gamma\partial_t+S^{(1)}_{r\theta\phi}
\end{equation}
with $\beta, \gamma, S^{(1)}_{r\theta\phi}$ purely spatial differential operators or functions, one has
\begin{equation}
    \mathcal{S}_{(1)}(t \psinzero)-t\mathcal{S}_{(1)}\psinzero=2 \beta \partial_t \psinzero + \gamma \psinzero.
\end{equation}
This term makes the integral time independent after multiplying by $\ps \psinzero$.
In passing, we also note that by plugging the ansatz \eqref{eq: ansatz psione} into $\projnone$, it is possible to use \eqref{eq: the product in coordinate} to determine the coefficient $c$ in \eqref{eq: mode shift onto same mode}. One has:
\begin{eqnarray}
  c &=& \langle \! \langle \psinzero, e^{-i \omegan^{(0)}t}\chi^{(1)}_{\boldsymbol n} \rangle \! \rangle \nonumber \\
  && +8 \pi i \omegan^{(1)} M^{4/3} \int\mathrm{d}r\mathrm{d}\theta \sin\theta \Delta^{-3}\Lambda R_{\boldsymbol n}^{(0)2}S_{\boldsymbol n}^{(0)2}.
\end{eqnarray}
Using these results, and assuming for concreteness that \eqref{eq: decomposition of S1} holds, we can write the second order frequency shift as:
\begin{eqnarray}
\label{eq: appendix - effective second order freq shift formula}
  && \normQNM \omegan^{(2)} = i \surf t^{a} \ps \psinzero \mathcal{S}_{(2)} \psinzero \nonumber \\
  &&+ 2 i \surf t^{a} \ps \psinzero \mathcal{S}_{(1)} (e^{-i \omegan^{(0)}t}\chi^{(1)}_{\boldsymbol n}) \nonumber \\
  && -2 \omegan^{(1)} \langle \! \langle \psinzero, e^{-i \omegan^{(0)}t}\chi^{(1)}_{\boldsymbol n} \rangle \! \rangle \nonumber \\
  && -2 \omegan^{(1)2}i \surf t^{a} \ps \psinzero \frac{\Lambda}{\Sigma \Delta} \psinzero \nonumber \\
  && +2 \omegan^{(1)} \surf t^{a} \ps \psinzero (2 \beta \partial_t + \gamma) \psinzero.
\end{eqnarray}

\section{Green function and integral representation of the mode shift}
\label{app: integral representation of mode shift}
In any given coordinate system, the Green function of the Teukolsky operator is defined by the equation
\begin{equation}
    \odagg G(x^a,x'^{a}) = \frac{\delta^{(4)}(x^a-x'^{a})}{\sqrt{-g}}.
\end{equation}
For the Teukolsky equation on a Kerr spacetime and using BL coordinates, we can write
\begin{equation}
    G(x^a,x'^{a}) = \frac{1}{2 \pi} \int_{-\infty+i c}^{\infty + i c} \mathrm{d}\omega e^{-i \omega(t-t')}g(\omega,\boldsymbol{x},\boldsymbol{x}'),
\end{equation}
where $c \ll 1$, $\boldsymbol{x}\equiv \{r,\Omega\}\equiv\{r,\theta,\phi\}$, and with $g(\omega,\boldsymbol{x},\boldsymbol{x}')$ solution of
\begin{equation}
    \hodagg_{\omega} g(\omega,\boldsymbol{x}, \boldsymbol{x}') = \frac{\delta(r-r')\delta(\theta-\theta')\delta(\phi-\phi')}{\Sigma \sin\theta},
\end{equation}
where we have used $\sqrt{-g} = \Sigma \sin\theta$. Recalling that $\Sigma \odagg = T$, where $T$ is the Teukolsky operator of the Kerr spacetime as written in BL coordinates (see e.g. Eq.~(4.7) of \cite{Teukolsky:1973ha}), we have the equation
\begin{equation}
    \hat T_{\omega} g(\omega,\boldsymbol{x}, \boldsymbol{x}') = \delta(r-r')\delta(\cos\theta-\cos\theta')\delta(\phi-\phi').
\end{equation}
We can further write $g(\omega,\boldsymbol{x}, \boldsymbol{x}')$ as
\begin{equation}
\label{eq: angular decomposition of GF}
    g(\omega,\boldsymbol{x}, \boldsymbol{x}') = \sum_{m \in \mathbb{Z}} \sum_{l \ge l_{min}} g^{\Omega}_{ l m}(\omega,\Omega,\Omega') g_{lm}(\omega,r,r')
\end{equation}
where $l_{min}=\mathrm{max}(2,|m|)$ and where we have defined
\begin{equation}
\label{eq: angular GF}
   g^{\Omega}_{ l m}(\omega,\Omega,\Omega') = e^{i m (\phi - \phi')} S_{lm\omega}(\theta)S_{lm\omega}(\theta'),
\end{equation}
with $S_{lm\omega}$ being a shorthand for $S_{slm\omega}|_{s=-2}$. The remaining $g_{lm}(\omega,r,r')$ then satisfies
\begin{equation}
    \hat T_{\omega m r}g_{lm}(\omega,r,r') = \delta(r-r'),
\end{equation}
with $- \hat T_{\omega m r}$ equal to the radial Teukolsky operator (see e.g. Eq.~(4.9) of \cite{Teukolsky:1973ha}). As this is simply an ODE, we have
\begin{equation}
    g_{lm}(\omega,r,r') = -\Delta^{-2}(r')\frac{R^{in}(r_<)R^{up}(r_{>})}{W_{lm}(\omega)}.
\end{equation}
With the Green function at hand, we can write the solution to 
\begin{equation}
    \hat T_{\omega}\chi(r,\theta,\phi) = Q(r,\theta,\phi)
\end{equation}
with vanishing initial data, as
\begin{equation}
    \chi(\boldsymbol{x}) = \int_V \mathrm{d}^3\boldsymbol{x}' g(\omega,\boldsymbol{x}, \boldsymbol{x}') Q(\boldsymbol{x}'),
\end{equation}
where the relevant volume element is $\mathrm{d}^3\boldsymbol{x} = \mathrm{d}r \mathrm{d}\theta \mathrm{d}\phi \sin\theta$ and where $V \equiv \{r\in [r_+,\infty), \theta \in [0,\pi), \phi \in (0,2\pi)\}$.

Consider now the equation satisfied by $\chi^{(1)}_{\boldsymbol n}$, the spatial part of the mode shift. It reads
\begin{equation}
  \hat  T_{\omegan^{(0)}}\chi^{(1)}_{\boldsymbol n} = -\left(\Sigma \hat{\mathcal{S}}^{(1)}_{\omegan^{(0)}} + \omegan^{(1)} \partial_{\omega} \hat T\big|_{\omegan^{(0)}} \right)\chi^{(0)}_{\boldsymbol n}.
\end{equation}
A formal solution then is 
\begin{eqnarray}
  \chi^{(1)}_{\boldsymbol n} &=& - \int \mathrm{d}^3\boldsymbol{x}'   g(\omegan^{(0)},\boldsymbol{x}, \boldsymbol{x}') \nonumber \\
  && \cdot \left(\Sigma' \hat{ \mathcal{S}}^{(1)}_{\omegan^{(0)}} + \omegan^{(1)} \partial_{\omega} \hat T\big|_{\omegan^{(0)}} \right) \chi^{(0)}_{\boldsymbol n}(\boldsymbol{x}').
\end{eqnarray}
At this level this solution is pathological. Indeed, looking at \eqref{eq: angular decomposition of GF}, we see that  $g(\omegan^{(0)},\boldsymbol{x}, \boldsymbol{x}')$ can be split in a regular part $g_{reg}$, featuring a sum 
over $(l',m')\neq(l,m)$ and a singular part $\tilde g$ with $l'=l,m'=m$. 
This piece is singular as $W_{lm}(\omegan^{(0)})=0$. Nevertheless, we proceed to show that $\tilde \chi^{(1)}_{\boldsymbol n}$ associated with the singular part $\tilde g$ is regular at $\omegan^{(0)}$. Indeed one has
\begin{eqnarray}
  \tilde \chi^{(1)}_{\boldsymbol n} &=& \frac{A_{\boldsymbol n}\chi^{(0)}_{\boldsymbol n}}{4 M^{4/3}W_{lm}(\omegan^{(0)})}\left( \surf t^a \ps \psinzero \mathcal{S}^{(1)}\psinzero  \right. \nonumber \\
  && + \left. \surf t^a \ps \psinzero \odaggzero(-i \omegan^{(1)}t \psinzero)\right),
\end{eqnarray}
and 
\begin{eqnarray}
  \surf t^a \ps \psinzero \mathcal{S}^{(1)}\psinzero &=& -i \normQNM \omegan^{(1)}, \nonumber \\
  \surf t^a \ps \psinzero \odaggzero(-i \omegan^{(1)}t \psinzero) &=& i \normQNM \omegan^{(1)}, \nonumber
\end{eqnarray}
where the first identity follows from \eqref{first order frequency shift}, and the second one can be shown to hold explicitly in BL coordinates. Hence, as currently written, $\tilde \chi^{(1)}_{\boldsymbol n}$ is undetermined. 
Defining $\chi^{(1)}_{\omega}$, with $\omega$ not a QNM frequency, as the solution of 
\begin{equation}
    \hat T_{\omega}\chi^{(1)}_{\omega} = -\left(\Sigma \hat{\mathcal{S}}^{(1)}_{\omegan^{(0)}} + \omegan^{(1)} \partial_{\omega} \hat T\big|_{\omegan^{(0)}} \right)\chi^{(0)}_{\boldsymbol n},
\end{equation}
we can write $\chi^{(1)}_{\omega}$ in terms of $g(\omega,\boldsymbol{x}, \boldsymbol{x}')$. Taking the limit $\omega \rightarrow \omegan^{(0)}$ we can exploit l'Hôpital's rule and relate the derivative of the Wronskian at a quasinormal frequency to the norm of the respective QNM to write
\begin{widetext}
    \begin{eqnarray}
\label{eq: integral representation of chi1 - regularized singular part}
  \tilde\chi^{(1)}_{\boldsymbol n} &=& \frac{i A_{\boldsymbol n}^{-1}4M^{4/3}}{\normQNM} \int \mathrm{d}^{3}\boldsymbol{x}' \Delta'^{-2} e^{i m (\phi - \phi')} \cdot \left. \frac{\mathrm{d}}{\mathrm{d}\omega}\left(S_{lm\omega}(\theta)S_{lm\omega}(\theta')R^{in}_{lm\omega}(r_<)R^{up}_{lm\omega}(r_>) \right) \right|_{\omegan^{(0)}} \nonumber \\
  && \times \left(\Sigma' \hat{\mathcal{S}}^{(1)}_{\omegan^{(0)}} + \omegan^{(1)} \partial_{\omega} \hat T\big|_{\omegan^{(0)}} \right)\chi^{(0)}_{\boldsymbol n}(\boldsymbol{x}').
\end{eqnarray}
The regular, integral representation of the spatial mode shift $\chi^{(1)}_{\boldsymbol n}$ is
\begin{equation}
    \label{eq: integral representation of chi1}
    \chi^{(1)}_{\boldsymbol n} = \tilde\chi^{(1)}_{\boldsymbol n} + \chi^{(1)}_{\boldsymbol n,reg},
\end{equation}
where
\begin{eqnarray}
\label{eq: integral representation of chi1 - regular part}
  \chi^{(1)}_{\boldsymbol n,reg} &=& -\int \mathrm{d}^3 \boldsymbol{x}' g_{reg}(\omegan^{(0)},\boldsymbol{x},\boldsymbol{x}') \left(\Sigma' \hat{\mathcal{S}}^{(1)}_{\omegan^{(0)}} + \omegan^{(1)} \partial_{\omega} \hat T\big|_{\omegan^{(0)}} \right)\chi^{(0)}_{\boldsymbol n}(\boldsymbol{x}').
\end{eqnarray}
\end{widetext}

\bibliography{refs}
\end{document}